\documentclass[12pt]{article}

\usepackage{epsfig}
\include{amssym}

\def\mathbf{\vec}

\def\ca{\c{c}\~{a}}

\def\ii{\'{\i}}

\newcommand{\njl}{\mathrm{NJL}}

\newcommand{\A}{\mathrm{A}}
\newcommand{\h}{\mathrm{H}}
\newcommand{\e}{\mathrm{E}}
\newcommand{\st}{\mathrm{st}}
\newcommand{\pt}{\mathrm{pt}}
\newcommand{\eff}{\mathrm{eff}}
\newcommand{\ud}{\mathrm{d}}

\begin{document}

\centerline{\Large\bf  
           Aspects of $U_\A (1)$ breaking }
\vspace{0.2cm}
\centerline{\Large\bf in the Nambu and Jona-Lasinio model}
\vspace{1cm}

\centerline{\large A. A. Osipov$^\dagger$\footnote{On leave from Joint 
            Institute for Nuclear Research, Laboratory of Nuclear 
            Problems, 141980 Dubna, Moscow Region, Russia. 
            Email address: osipov@nusun.jinr.ru}, 
            B. Hiller$^\dagger$\footnote{Email address: 
            brigitte@teor.fis.uc.pt},
            V. Bernard$^\ddagger$\footnote{
            Email address: bernard@lpt6.u-strasbg.fr},
            A. H. Blin$^\dagger$\footnote{Email address:
            alex@teor.fis.uc.pt}} 
\vspace{0.5cm}
\centerline{$^\dagger$\small\it Centro de F\'{\i}sica Te\'{o}rica, 
            Departamento de
            F\'{\i}sica da Universidade de Coimbra,}
\centerline{\small\it 3004-516 Coimbra, Portugal}
\vspace{0.2cm}
\centerline{$^\ddagger$\small\it Universit\'e Louis Pasteur, 
            Laboratoire de Physique Th\'eorique 3-5,}  
\centerline{\small\it rue de l'Universit\'e, F-67084 Strasbourg, France}
\vspace{1cm}

\centerline{\bf Abstract}
\vspace{0.5cm}
The six-quark instanton induced 't Hooft interaction, which breaks 
the unwanted $U_\A (1)$ symmetry of QCD, is a source of
perturbative corrections to the leading order result formed by the 
four-quark forces with the $U_L (3)\times U_R (3)$ chiral symmetry.
A detailed quantitative calculation is carried out to bosonize the 
model by the functional integral method. We concentrate our efforts
on finding ways to integrate out the auxiliary bosonic variables.
The functional integral over these variables cannot be evaluated 
exactly. We show that the modified stationary phase approach leads
to a resummation within the perturbative series and calculate the
integral in the ``two-loop'' approximation. The result is a correction 
to the effective mesonic Lagrangian which may be important for the 
low-energy spectrum and dynamics of the scalar and pseudoscalar nonets. 

\vspace{1.0cm}
PACS number(s): 12.39.Fe, 11.30.Rd, 11.30.Qc

\newpage 

\section*{\centerline{\large\sc 1. Introduction}}

In the absence of a quantitative framework within QCD to deal 
with its large distance dynamics, the physics of hadrons is usually 
approached either through an effective field theory or through 
phenomenological parametrizations based on some simple ansatz with 
solid symmetry grounds. In the first case the theory is written in
terms of mesonic degrees of freedom \cite{Weinberg:1979,Gasser:1983}. 
In the second case the existence of underlying multi-quark
interactions can be assumed. In this way one takes into account the 
quark structure of mesons explicitly, which may yield very important 
information about the structure of the QCD vacuum. Although these 
interactions are not renormalizable (as compared with CHPT, which is 
renormalizable in the sense of effective field theories), it is often 
possible to derive useful results in such models by introducing an 
ultraviolet cutoff. 

Not much is known about the origin of multi-quark vertices. The 
semi-classical theory based on the QCD instanton vacuum 
\cite{Diakonov:2003} provides evidences in favor of these
interactions: two or more quarks can scatter off the same instanton 
(or anti-instanton), certain correlations between quarks originate 
from averaging over their positions and orientations in color space, 
the result being an effective quark Lagrangian. Assuming a dominant 
role for quark zero modes in this scattering process one obtains 
$2N_f$-quark interactions ($N_f$ is the number of quark flavors), 
which are known as 't Hooft interactions \cite{Hooft:1978}. Actually, 
an infinite number of multi-quark interactions, starting from the 
four-quark ones, has been found in the instanton-gas model considered 
beyond the zero mode approximation, all of them being of the same 
importance \cite{Simonov:1997}. 

On the other hand, accurate lattice measurements for the realistic QCD 
vacuum show a hierarchy between the gluon field correlators with a 
dominance of the lowest one \cite{Bali:2001}. A similar hierarchy of 
the multi-quark interactions can be triggered after averaging over 
gluon fields. If this is true, there is an apparent contradiction with 
the instanton-gas model. This point has been stressed in 
\cite{Simonov:2002}.

The hierarchy problem of multi-quark interactions can be addressed 
on pure phenomenological grounds. For that we suggest to simplify the 
task considering only the four- and six-quark interactions. 
Once one knows the Lagrangian the obvious question arises: does the
system possess a stable vacuum state and does this state correspond
to our phenomenological expectations? If hierarchy takes place this
question is pertinent for the leading four-quark interaction, 
because in this case the effective quark Lagrangian can be studied 
step by step in the hierarchy with the assumption that four-quark 
vertices are the most important ones. In the opposite case we must 
study the system as a whole to answer the question. For the best 
known and simplest example, to which we dedicate most of our 
attention here, the solution may be found analytically. As a result 
one can obtain definite answers to the above questions with a 
convincing indication in favor of a hierarchy for the considered 
example.   

Our choice of model is not accidental. The importance of the 
four-fermion interactions has been recognized for many years, starting in 
the early sixties, when Nambu and Jona-Lasinio (NJL) \cite{Nambu:1961} 
used it for studying dynamical breaking of chiral symmetry. Later on 
a modified form of this interaction, written in terms of quark fields, 
has been used to derive the QCD effective action at long distances 
\cite{Eguchi:1976,Volkov:1982,Ebert:1986}. 

The six-quark vertices contain additional information about the 
vacuum \cite{Dorokhov:1992}.
They break explicitly the $U_\A (1)$ symmetry and are the only source 
of OZI-violating effects \cite{Okubo:1963}. Taken together with the NJL 
interactions, the 't Hooft Lagrangian gives a good description of the 
pseudoscalar nonet, especially the $\eta$ and $\eta'$ masses and mixing 
\cite{Bernard:1988,Reinhardt:1988}. In this form the model has been 
widely explored at the mean-field level \cite{Hatsuda:1994}. 

Let us discuss the Lagrangian which we shall use in our analysis.
On lines suggested by multicolor chromodynamics it can be argued 
\cite{Witten:1979a} that the $U_\A (1)$ anomaly vanishes in the 
large $N_c$ limit, so that mesons come degenerate in mass nonets. 
Hence the leading order (in $N_c$ counting) mesonic Lagrangian and 
the corresponding underlying quark Lagrangian must inherit the 
$U_L(3)\times U_R(3)$ chiral symmetry of massless QCD. In accordance
with these expectations the $U_L(3)\times U_R(3)$ symmetric NJL 
interactions, 
\begin{equation}
\label{L4q}
  {\cal L}_\njl =\frac{G}{2}\left[(\bar{q}\lambda_aq)^2+
                 (\bar{q}i\gamma_5\lambda_aq)^2\right],
\end{equation}
can be used to specify the corresponding local part of the effective 
quark Lagrangian in channels with quantum numbers $J^P=0^+, 0^-$.
The Gell-Mann matrices acting in flavor space, $\lambda_a,\
a=0,1,\ldots ,8,$ obey the basic property $\mbox{tr} \lambda_a
\lambda_b =2\delta_{ab}$. 

The 't Hooft determinantal interactions are described by the 
Lagrangian \cite{Hooft:1978} 
\begin{equation}
\label{Ldet}
  {\cal L}_{\h}=\kappa (\mbox{det}\ \bar{q}P_Lq
                         +\mbox{det}\ \bar{q}P_Rq)
\end{equation}
where the matrices $P_{L,R}=(1\mp\gamma_5)/2$ are projectors and the 
determinant is over flavor indices. 

The coupling constant $\kappa$ is a dimensional parameter 
($[\kappa ]=\mbox{GeV}^{-5}$) with the large $N_c$ asymptotics 
$\kappa\sim 1/N_c^{N_f}$. The coupling $G$, $[G]=\mbox{GeV}^{-2}$, 
counts as $G\sim 1/N_c$ and, therefore, the Lagrangian (\ref{L4q}) 
dominates over ${\cal L}_\h$ at large $N_c$. It differs from the 
counting $G\sim 1/N_c^2$, which one obtains in the instanton-gas 
vacuum \cite{Simonov:1997}. 

It is assumed here for simplicity that interactions between quarks 
can be taken in the long wavelength limit where they are effectively 
local. The 't Hooft-type ansatz (\ref{Ldet}) is a frequently used 
approximation. Even in this essentially simplified form the 
determinantal interaction has all basic ingredients to describe the 
dynamical symmetry breaking of the hadronic vacuum and explicitly 
breaks the axial $U_\A (1)$ symmetry \cite{Diakonov:1996}. The 
effective mesonic Lagrangian, corresponding to the non-local 
determinantal interaction, has been found in \cite{Diakonov:1985}.   

Anticipating our result, we would like to note that if the hierarchy 
of multi-quark interactions really occurs in nature, the perturbative 
treatment seems adequate. The NJL interaction alone has a stable 
vacuum state corresponding to spontaneously broken chiral symmetry. 
But, as we shall show, the effective quark theory based on the 
Lagrangian\footnote{The other multi-quark terms have been 
neglected here.} 
\begin{equation}
\label{totlag}
  {\cal L}=\bar{q}(i\gamma^\mu\partial_\mu -\hat{m})q
          +{\cal L}_\njl + {\cal L}_\h
\end{equation}
has a fatal flaw: if ${\cal L}_\h$ is comparable with ${\cal L}_\njl$, 
it has no stable ground state. This feature of the model is invisible 
in a perturbative approach in ${\cal L}_\h$. The situation is exactly 
analogous to the problem of a harmonic oscillator perturbed by an
$x^3$ term. This system has no ground state, but perturbation theory 
around a local minimum does not know this. In eq.(\ref{totlag}) the 
current quark mass, $\hat{m}$, is a diagonal matrix with elements 
$\mbox{diag} (\hat{m}_u, \hat{m}_d, \hat{m}_s)$, which explicitly 
breaks the global chiral $SU_L(3)\times SU_R(3)$ symmetry of the 
Lagrangian.

There is also a special problem related with the bosonization of multi-quark 
interactions. To bosonize the theory one introduces  
auxiliary bosonic variables to render fermionic vertices bilinear
in the quark fields. This procedure requires twice more bosonic 
degrees of freedom than necessary \cite{Reinhardt:1988}. 
Redundant variables must be integrated out and this integration is 
problematic as soon as one goes beyond the lowest order stationary
phase approximation \cite{Osipov:2002,Osipov:2004}: the lowest 
order result is simply the value of the integrand taken at one
definite stationary point \cite{Diakonov:1996}. In this paper we shall 
show for the first time how to extract
systematically the higher order corrections which contribute to the 
effective mesonic Lagrangian, and what to do with infinities
contained in these corrections. The model (\ref{totlag}) is considered to  
illustrate our calculations. If the coupling $\kappa$ is not too 
small, corrections can be much larger than one might expect from
$1/N_c$ counting, and be important for the mesonic ($0^+,0^-$) mass 
spectra and their dynamics. The large value of the $\eta - \eta'$ 
mass difference obtained already at leading order and totally
determined by the six-quark interactions gives some credit to large
corrections. Additionally, one can expect some enhancement at next
to leading order by virtue of divergent factors: the cutoff scale 
is not known beforehand, and can be relatively large.     

Our study represents a very simplified view on the matter and should
be considered as a first rough estimate which can be improved if 
necessary. The most essential approximation made here is 
related with the local character of the multi-quark vertices in 
eq.(\ref{totlag}). It leads to $\delta (0)\,$-singularities beyond the 
mean-field framework (see also \cite{Osipov:2002,Osipov:2004}).  
Nevertheless, if it turns out that the low-energy QCD vacuum contains 
the hierarchy of multi-quark interactions, our approach can be used 
as a basis for a more serious work in this direction.       

The paper is organized as follows. In Section 2 the functional integral
representation, ${\cal Z}$, for a bosonized version of the NJL model 
with 't Hooft interactions is derived. To make clear the approximations 
which will be used for the functional integration, a similar non 
functional integral, $I$, is considered in Section 3. The reader who 
just wants to get a general idea of what we intend to study, can find
it here. This section contains several clarifying points which are 
important in the following. In subsection 3.1 we give the exact result 
for $I$. Its stationary phase asymptotics is obtained in subsection 3.2. 
The semi-classical method yields the same result, as shown in
subsection 3.3, as does the method presented in subsection 3.4. The
point of these numerous calculations is just to show that all methods
considered lead to the same asymptotics for the integral $I$ which is totally 
determined by the number of stationary points. The perturbative 
treatment is given in subsection 3.5. Here we discuss why the perturbation
theory result is essentially different from the asymptotics obtained in
the previous subsections, even for very small value of the expansion 
parameter. We resum the perturbative series in subsection 3.6. Calculating 
the next to leading order correction we illustrate the main goal of our 
studies in the forthcoming sections. 

In Section 4 (subsection 4.1) we show that the chiral symmetry group 
imposes strong constraints which are only compatible either with
the perturbative approach, or the expansion in a parameter that 
multiplies the total Lagrangian density (the loop expansion). Otherwise, 
as the consistent stationary phase treatment shows (subsection 4.2), 
the model is unstable. 

The first alternative is considered in Section 5. To evaluate the  
functional integral ${\cal Z}$, we consider it like the natural
infinite-dimensional limit of ordinary finite-dimensional Gaussian
integrals (subsection 5.1). The perturbative treatment essentially 
simplifies calculations. Nevertheless, the integration over auxiliary 
variables leads to a special problem with $\delta (0)\,$-singularities. 
We discuss this aspect of the bosonization in subsection 5.2. Another 
way to obtain the result is shown in subsection 5.3. Here a more 
elaborate spectral representation method is used to justify our 
computations. This subsection contains the prescriptions for the 
regularization of the $\delta (0)$ infinities, and a discussion of 
their reliability. 
 
The second alternative (the loop expansion) is considered in Section 6.  
Here, in subsection 6.1, we obtain in closed form the two-loop 
contributions to the functional integral ${\cal Z}$ and give arguments
to justify this result. In subsection 6.2 we end with short conclusions 
and suggest future applications of our result. 
 
The summary is given and outlook is surveyed in Section 7.

The Appendix contains the definition and the main properties 
of Airy's functions.

\vspace{0.7cm}

\section*{\centerline{\large\sc 2. Bosonization}}

The many-fermion vertices of Lagrangian (\ref{totlag}) can be 
linearized by introducing the functional unity \cite{Reinhardt:1988}
\begin{eqnarray}
\label{1}
  1&\!\!\! =\!\!\!& \int \prod_a {\cal D}s_a{\cal D}p_a\ \delta 
                    (s_a-\bar{q}\lambda_aq) 
                    \delta (p_a-\bar{q}i\gamma_5\lambda_aq)
                    \nonumber \\
   &\!\!\! =\!\!\!& \int \prod_a {\cal D}s_a {\cal D}p_a
                    {\cal D}\sigma_a {\cal D}\phi_a
                    \nonumber \\  
   &\!\!\!\times \!\!\!& \exp \left\{i\int\!\ud^4x 
                    [\sigma_a(s_a-\bar{q}\lambda_aq) +
                     \phi_a(p_a-\bar{q}i\gamma_5\lambda_aq)]\right\}
\end{eqnarray}
in the vacuum-to-vacuum amplitude  
\begin{equation}
\label{genf1}
   Z=\int {\cal D}q{\cal D}\bar{q}\ \exp\left(i\int\! \ud^4x{\cal L}\right).
\end{equation}  
We consider the theory of quark fields in four-dimensional Minkowski 
space. It is assumed that the quark fields have color $(N_c=3)$ and flavor 
$(N_f=3)$ indices which range over the set $i=1,2,3$. 
The auxiliary bosonic fields, $\sigma_a$, and, $\phi_a, (a=0,1,\ldots,
8)$ become the composite scalar and pseudoscalar mesons and the auxiliary 
fields, $s_a$, and, $p_a$, must be integrated out. 

By means of the simple trick (\ref{1}), it is easy to write down the 
amplitude 
(\ref{genf1}) as 
\begin{equation}
\label{genf2}
   Z=\int {\cal D}q{\cal D}\bar{q}
     \prod^8_{a=0}{\cal D}s_a
     \prod^8_{a=0}{\cal D}p_a
     \prod^8_{a=0}{\cal D}\sigma_a
     \prod^8_{a=0}{\cal D}\phi_a
     \exp\left(i\int \ud^4x{\cal L}'\right)
\end{equation}  
with
\begin{equation}
\label{sembosL}
      {\cal L}'= \bar{q}(i\gamma^\mu\partial_\mu -\sigma 
               - i\gamma_5\phi )q + s_a (\sigma_a -\hat{m}_a)
               + p_a\phi_a + {\cal L}_\njl' + {\cal L}_\h'\ ,
\end{equation}
\begin{equation}
\label{lagrNJL}
      {\cal L}_\njl' = \frac{G}{2}\left[(s_a)^2+(p_a)^2\right]\ ,
\end{equation}
\begin{equation}
      {\cal L}_\h' = \frac{\kappa}{64}\left[ \mbox{det}(s+ip)
                     +\mbox{det}(s-ip)\right] 
                   = \frac{\kappa}{32}A_{abc}s_a
                     \left(s_bs_c-3p_bp_c \right).
\end{equation}
We assume here that $\sigma =\sigma_a\lambda_a$, and so on for all
auxiliary fields $\sigma ,\phi ,s,p$. The totally symmetric constants 
$A_{abc}$ are related to the flavor determinant, and equal to
\begin{eqnarray}
\label{A}
   A_{abc}\!\!\!\!\!\!\!\!
   &&=\frac{1}{3!}\epsilon_{ijk}\epsilon_{mnl}(\lambda_a)_{im}
             (\lambda_b)_{jn}(\lambda_c)_{kl} \nonumber \\
   &&=\frac{2}{3}d_{abc} +
      \sqrt{\frac{2}{3}} \Big(
      3\delta_{a0}\delta_{b0}\delta_{c0}
      -\delta_{a0}\delta_{bc}
      -\delta_{b0}\delta_{ac}
      -\delta_{c0}\delta_{ab}\Big).
\end{eqnarray}
We use the standard definitions for antisymmetric $f_{abc}$ and
symmetric $d_{abc}$ structure constants of $U(3)$ flavor symmetry. 
One can find, for instance, the following useful relations
\begin{eqnarray}
   &&f_{eac}A_{bfc}+f_{ebc}A_{fac}+f_{efc}A_{abc}=0, \nonumber \\ 
   &&d_{eac}A_{bfc}+d_{ebc}A_{fac}+d_{efc}A_{abc} 
     = \sqrt{6}\delta_{e0}A_{abf}, \nonumber \\
   &&\sum_{b=0}^8 A_{abb} = -2 \sqrt{\frac{2}{3}}\ \delta_{a0},
     \qquad \sum_{c,e=0}^{8} A_{ace}A_{bce} = \frac{8}{9}\ \delta_{ab}. 
\end{eqnarray}

At this stage it is easy to rewrite eq.(\ref{genf2}), by changing 
the order of integrations, in a form appropriate to accomplish the 
bosonization, i.e., to calculate the integrals over quark fields 
and integrate out from $Z$ the unphysical part associated with the 
auxiliary bosonic variables ($s_a,\ p_a$) 
\begin{eqnarray}
\label{genf3}
   Z&\!\!\! =\!\!\!&\int \prod_a{\cal D}\sigma_a{\cal D}\phi_a
                    {\cal D}q{\cal D}\bar{q}\
                    \exp\left(i\int\ud^4x
                    {\cal L}_q(\bar{q},q,\sigma ,\phi )\right)
                    \nonumber \\
    &\!\!\!\times\!\!\!&\int \prod_a{\cal D}s_a{\cal D}p_a\
    \exp\left(i\int\ud^4x{\cal L}_r(\sigma ,\phi ,s,p)\right)
\end{eqnarray}  
where
\begin{eqnarray}
\label{lagr2}
  {\cal L}_q&\!\!\! =\!\!\!&
  \bar{q}(i\gamma^\mu\partial_\mu -\sigma - i\gamma_5\phi )q, \\
\label{lagr3}
  {\cal L}_r&\!\!\! =\!\!\!& s_a(\sigma_a - \hat{m}_a) + p_a\phi_a
  + {\cal L}'_\njl + {\cal L}'_\h\ .            
\end{eqnarray}
The Fermi fields enter the action bilinearly, thus one can always integrate
over them, since one deals with a Gaussian integral. One should also
shift the scalar fields $\sigma_a(x)\rightarrow\sigma_a(x)+m_a$ by 
demanding that the vacuum expectation values of the shifted fields 
vanish $\big < 0|\sigma_a(x)|0\big >=0$. In other words, all tadpole 
graphs in the end should sum to zero, giving us the gap equation to fix the 
constituent quark masses $m_a$ corresponding to the physical vacuum state.

The functional integrals over $s_a$ and $p_a$  
\begin{equation}
\label{intJ}
     {\cal Z}[\sigma ,\phi ;\Delta ]\equiv 
     {\cal N}\int\limits^{+\infty}_{-\infty}\prod_a{\cal D}s_a{\cal D}p_a\
     \exp\left(i\int\ud^4x{\cal L}_r(\sigma +m,\phi ,s,p)\right)
\end{equation} 
are the main subject of our study. We put here $\Delta_a=m_a-\hat{m}_a$, 
and ${\cal N}$ is chosen so that ${\cal Z}[0,0;\Delta ]=1$. 

Let us join the auxiliary bosonic variables in one $18$-component object 
$R_A=(R_a,R_{\dot{a}})$ where we identify $R_a\equiv s_a$ and
$R_{\dot{a}}\equiv p_a$; $a,\dot{a}$ run from $0$ to $8$
independently. 
It is clear then, that $R_A^2=s_a^2+p_a^2$.
Analogously, we will use $\Pi_A=(\sigma_a,\phi_a)$ for external 
fields and $\Delta_A=(\Delta_a,0)$. 

Next, consider the sum $\Phi_{ABC}R_AR_BR_C$. If we require 
\begin{equation} 
   \Phi_{abc}=\frac{3}{16}A_{abc}\, , \quad 
   \Phi_{a\dot{b}\dot{c}}=-\frac{3}{16}A_{abc}\, , \quad 
   \Phi_{ab\dot{c}}=0, \quad 
   \Phi_{\dot{a}\dot{b}\dot{c}}=0, 
\end{equation}
we find after some algebra
\begin{equation} 
   \frac{\kappa}{3!}\ \Phi_{ABC}R_AR_BR_C = {\cal L}_\h'
\end{equation}
with the following important property to be fulfilled
\begin{equation}
\label{contr}
      \Phi_{ABC}\delta_{BC}=0.
\end{equation}

Now it is easy to see that the functional integral (\ref{intJ}) can be
written in a compact way
\begin{equation}
\label{intJi}
     {\cal Z}[\Pi ,\Delta ]\equiv 
     {\cal N}\int\limits^{+\infty}_{-\infty}\prod_A{\cal D}R_A\
     \exp\left(i\int\ud^4x{\cal L}_r(\Pi ,\Delta ;R)\right),
\end{equation} 
where
\begin{equation}
\label{Lr}
   {\cal L}_r = R_A (\Pi_A+\Delta_A) + \frac{G}{2} R_A^2 
   + \frac{\kappa}{3!}\ \Phi_{ABC}R_AR_BR_C.
\end{equation}
We have arrived at a functional integral with a cubic polynomial
in the exponent. 

\vspace{0.7cm}

\section*{\large\sc 3. Digression to a one dimensional case}

To get a rough idea of how to evaluate the integral (\ref{intJi}) 
we start with its one-dimensional analog
\begin{equation}
\label{I1}
   I = \int\limits_{-\infty}^{+\infty}\! \ud x\ e^{iP(x)}
\end{equation}
where
\begin{equation}
   P(x) = \sigma x + \frac{a}{2!}x^2 + \frac{b}{3!} x^3\, ,  
\end{equation}
and $\sigma ,a,b$ are constants. This integral plays the same role as
our desired functional one, but is well defined as an improper
Riemann integral of the real variable $x$.  

\subsection*{\normalsize\it 3.1 The exact result}
The integral (\ref{I1}) can be evaluated exactly. To show this let us 
express the polynomial $P(x)$ in the form  
\begin{equation}
   P(x) = P(x_0) + P'(x_0)(x-x_0)
                 + \frac{P'''(x_0)}{3!}(x-x_0)^3\, , 
\end{equation}
where $x_0$ is chosen to satisfy the following equation
\begin{equation}
\label{cpeq}
   P''(x)= bx +a = 0 
   \ \ \ \Rightarrow\ \ \  
   x_{0} = -\frac{a}{b}\, . 
\end{equation}
The coefficients are
\begin{equation}
\label{Px0}
   P(x_0) = -\frac{a\sigma}{b} + \frac{a^3}{3b^2}\, , \quad 
   P'(x_0) = \sigma - \frac{a^2}{2b}\, , \quad P'''(x)=b.
\end{equation}

Hence the integral (\ref{I1}) is given by
\begin{equation}
\label{I1a}
   I = e^{iP(x_0)} \int\limits_{-\infty}^{+\infty}\! \ud x\,
               \exp\left[ i \left( P'(x_0) x
               + \frac{P'''(x_0)}{3!} x^3 \right)\right]\, .
\end{equation}
We can rewrite eq.(\ref{I1a}) in terms of the new variable $t$
\begin{equation}
\label{repl1}
   x = \left(\frac{2}{|P'''(x_0)|}\right)^{1/3} t\, ,
\end{equation}
thus we arrive at
\begin{equation}
\label{sp1}
   I = e^{iP(x_0)}\left(\frac{2}{|b|}\right)^{1/3}
       \int\limits_{-\infty}^{+\infty}\! \ud t\,
       \exp\left( i \left( \frac{t^3}{3}+\delta t\right)\right)\, ,
\end{equation}
where $\delta$ is defined to be  
\begin{equation}
   \delta = \mbox{sgn}(b)\, P'(x_0)\left(\frac{2}{|b|}
             \right)^{1/3}.
\end{equation}
This integral is well known. The result of integration can be 
represented in terms of the Airy function (see Appendix for details)
\begin{equation}
\label{I-exact}
   I = 2\pi\, e^{iP(x_0)}\left(\frac{2}{|b|}\right)^{1/3} Ai(\delta ).
\end{equation}
If $\delta$ is real, the function $Ai(\delta )$ is also real, and
the phase of $I$ is equal to $P(x_0)$. In the following, after some
generalizations made for the integral $I$, its phase will represent  
the effective action of a dynamical system. 
The expression for this phase is the main goal of our calculations. 

\subsection*{\normalsize\it 3.2 The stationary phase result}
The result (\ref{I-exact}) for $I$ is exact. It can be approximated 
at large values of $\delta$ by its asymptotic series, still giving 
us an exact expression for the phase. To obtain asymptotics let us 
transform eq.(\ref{sp1}) to a more convenient form by the replacement
\begin{equation}
   t = \sqrt{|\delta|}\ y\, .
\end{equation} 
One has
\begin{equation}
\label{sp2}
   I = e^{iP(x_0)}\left(\frac{2}{|b|}\right)^{1/3}\sqrt{|\delta|}
       \int\limits_{-\infty}^{+\infty}\! \ud y\,
       \exp\left( i |\delta |^{3/2}g(y) \right), 
\end{equation}
where $g(y)=y^3/3+\mbox{sgn}(\delta )y$. The integral (\ref{sp2}) 
is already in the form appropriate for the stationary phase method 
to be applied. Indeed, as $\delta\rightarrow -\infty$ the term 
$|\delta |^{3/2} g(y)$ gives a rapidly oscillating contribution to 
the integrand in eq.(\ref{sp2}) which cancels out, except at the 
regions of critical points. The contribution from these regions can 
be evaluated on the basis of the stationary phase method. In our case 
$a\sim G\sim 1/N_c$, $b\sim\kappa\sim 1/N_c^3$, $\sigma\sim 1$, i.e., 
$\delta^{3/2}\sim N_c^3$ and large $N_c$ arguments\footnote{All large 
$N_c$ arguments given in this section hint of course at the functional
integral case which will be discussed later.} 
can be used to 
justify the $\delta\to -\infty$ limit. Both critical points 
($g'(y)=0 \Rightarrow y_{1,2}=\pm 1$) belong to the interval of 
integration. Thus we have
\begin{eqnarray}
   I&\!\!\! \sim \!\!\!& 
      e^{iP(x_0)}\left(\frac{2}{|b|}\right)^{1/3}\!\sqrt{|\delta|}
      \int\limits_{-\infty}^{+\infty}\! \ud y 
      \left\{ 
      \exp\left[ i |\delta |^{3/2} 
      \left(-\frac{2}{3}+y^2+\frac{y^3}{3}\right) \right]
      \nonumber \right.\\
    &\!\!\! +\!\!\! & \left.
      \exp\left[ i |\delta |^{3/2} 
      \left(\frac{2}{3}-y^2+\frac{y^3}{3}\right) \right]
      \right\} \qquad (\delta\to -\infty ).
\end{eqnarray}
The last term in both exponents can be factorized and then expanded
in a power series of $y^3$ 
\begin{eqnarray}
\label{ias}
   I&\!\!\! \sim \!\!\!& 
      e^{iP(x_0)}\left(\frac{2}{|b|}\right)^{1/3}\!\sqrt{|\delta|}
      \int\limits_{-\infty}^{+\infty}\! \ud y 
      \left\{ 
      \exp\left[ i |\delta |^{3/2} 
      \left(y^2-\frac{2}{3} \right) \right]
      \nonumber \right.\\
    &\!\!\! +\!\!\! & \left.
      \exp\left[ -i |\delta |^{3/2} 
      \left(y^2-\frac{2}{3} \right) \right]
      \right\}\sum_{n=0}^{\infty}\frac{1}{n!}
      \left(i|\delta |^{3/2}\frac{y^3}{3}\right)^n
      \quad (\delta\to -\infty ),
\end{eqnarray}
and integrated term by term. The corresponding integrals 
are evaluated exactly
\begin{equation}
   \int\limits_{-\infty}^{+\infty}\! \ud y\, \exp 
   \left(\pm i\alpha y^2 \right) y^{2n}
   = (\pm i)^n \frac{(2n-1)!!}{(2\alpha)^n}\
   e^{\pm i\pi /4} \left( \frac{\pi}{\alpha}\right)^{1/2}  
   \qquad (\alpha >0)
\end{equation}
giving as a result the following asymptotics for the integral $I$ 
\begin{equation}
\label{asympt}
   I \sim \frac{2^{4/3}\sqrt{\pi}}{|\delta |^{1/4}|b|^{1/3}}
          e^{iP(x_0)}
          \left[\cos\left(\frac{2}{3} |\delta |^{3/2}-\frac{\pi}{4}\right)
          +{\cal O}(|\delta |^{-3/2})\right]
\qquad (\delta\to -\infty ).
\end{equation} 
Let us stress that both critical points contribute to the resulting
asymptotic series reproducing the well known asymptotics of the Airy's
function (\ref{Airey-as2}). We have considered here the case when the series
oscillates. In the opposite case, $\delta\to\infty$, the asymptotics falls  
down exponentially (\ref{Airey-as1}), and does not have the appropriate 
form from the physical point of view. 

\subsection*{\normalsize\it 3.3 Semi-classical asymptotics}
One can calculate the integral (\ref{I1}) assuming that a large 
parameter is already present in the exponent. For instance, the 
reduced Planck constant, $\hbar $, can be considered as such a parameter 
\begin{equation}
   I = \int\limits_{-\infty}^{+\infty}\!\ud x\, \exp \left(
       \frac{i}{\hbar}\, P(x)\right).
\end{equation}
In this case one obtains the asymptotic expansion for $I$ at 
small values of $\hbar\to 0$. This approach is known as the 
semi-classical expansion. 

The real function $P(x)$ has two critical points $x_j\ (j=1,2)$  
\begin{equation}
\label{crit}
   P'(x)=\frac{b}{2}x^2+ax+\sigma =0 
   \ \ \ \Rightarrow\ \ \  
   x_{1,2} = \frac{-a\pm\sqrt{a^2-2b\sigma}}{b}\, . 
\end{equation}
Both of them are real at $a^2>2b\sigma$ and, therefore, belong 
to the contour of integration. In the neibourhood of these points the
polynomial $P(x)$ is conveniently written in the form
\begin{equation}
\label{RA}
   P(x) = P(x_j) + \frac{P''(x_j)}{2!}(x-x_j)^2
                 + \frac{P'''(x_j)}{3!}(x-x_j)^3\, , 
\end{equation}
where
\begin{eqnarray}
   P(x_j) &\!\!\! =\!\!\! & -\frac{a\sigma}{b} + \frac{a^3}{3b^2} 
            + (-1)^j\ \frac{D^{3/2}}{3b^2}\, , \quad
   D\equiv a^2-2b\sigma >0, \nonumber \\
   P''(x_j) &\!\!\! =\!\!\! & (-1)^{j+1}D^{1/2}, \quad P'''(x)=b.
\end{eqnarray}
Thus the integral under consideration is estimated as
\begin{equation}
\label{sem}
   I \sim \sum_{j=1}^{2} e^{iP(x_j)}
               \int\limits_{-\infty}^{+\infty}\!\ud x\,
               \exp\left[ i \left( \frac{P''(x_j)}{2} x^2
               + \frac{P'''(x_j)}{3!} x^3 \right)\right]
          \qquad (\hbar\to 0),
\end{equation}
or 
\begin{equation}
\label{ralk}
   I \sim \frac{2^{3/2}\sqrt{\pi}}{D^{1/4}}\ e^{iP(x_0)}
          \left[\cos\left(\frac{D^{3/2}}{3b^2}
                           -\frac{\pi}{4}\right)
          +{\cal O}(\hbar )\right]
          \qquad (\hbar\to 0).
\end{equation} 
Noting that
\begin{equation}
   D = -\delta \sqrt[3]{4b^4}\, ,
\end{equation}
one can see that this result coincides with our previous estimate 
(\ref{asympt}). 

It may be helpful to remark that if one would take the contribution 
of only one critical point in (\ref{sem}), the resulting asymptotics 
would be obviously different, and, what is important for us, the phase 
too. 

\subsection*{\normalsize\it 3.4 The $N_c\to\infty $ asymptotics}
The representation (\ref{RA}) can be used as a first step to 
estimate the integral $I$ without any reference to the 
semi-classical expansion. Alternatively, our calculations can be 
based on the large $N_c$ asymptotics. According to the formula
(\ref{RA}), we have identically
\begin{equation}
   I = e^{iP(x_j)} \int\limits_{-\infty}^{+\infty}\!\ud x\,
               \exp\left[ i \left( \frac{P''(x_j)}{2} x^2
               + \frac{P'''(x_j)}{3!} x^3 \right)\right]\, .
\end{equation}
This holds whether $x_1$ or $x_2$ (see eq.(\ref{crit})) is chosen 
here.

Next, replacing the variables
\begin{equation}
   x = \bigg{|}\frac{3P''(x_j)}{P'''(x_j)}\bigg{|}\, t\, ,
\end{equation}
we arrive at
\begin{equation}
\label{sp}
   I = e^{iP(x_j)}\frac{3\sqrt{D}}{|b|}
       \int\limits_{-\infty}^{+\infty}\!\ud t\,
       \exp\left( i \lambda f(t)\right)\, ,
\end{equation}
where we have used the following notations
\begin{equation}
   \lambda = \frac{9D^{3/2}}{2b^2}\, , \qquad
   f(t) = t^2\left[(-1)^{j+1} + \mbox{sgn}(b)\, t\right].
\end{equation}
This form of the integral is already appropriate to apply the 
stationary phase method at large $\lambda$. Large $N_c$ 
arguments can be used to justify the $\lambda\to\infty$ limit,
for it is known that $\lambda\sim N_c^3$. 

Let us obtain the leading term in this asymptotics. The critical points
are given by the equation
\begin{equation} 
   f'(t)=0\ \ \ \Rightarrow\ \ \ t_1=0, \ \ 
   t_2=(-1)^j\ \mbox{sgn}(b)\,\frac{2}{3}\, . 
\end{equation}
Therefore, the integral $I$ has the following asymptotical estimate 
at $\lambda\to\infty$
\begin{eqnarray}
   I&\!\!\!\!\sim \!\!\!\!& 
   e^{iP(x_j)}\frac{3\sqrt{D}}{|b|}
   \int\limits_{-\infty}^{+\infty}\!\ud t\, \left\{
   \exp\left[ i \lambda \left( (-1)^{j+1}t^2+
   \mbox{sgn}(b)\,t^3\right)\right] \right.\nonumber \\
   &\!\!\!\! + \!\!\!\!& \left.
   \exp\left[i\lambda\left((-1)^{j+1}\frac{4}{27}+
   (-1)^{j}t^2+\mbox{sgn}(b)\,t^3\right)\right]\right\}.
\end{eqnarray}
The first term of the integrand is the contribution of the critical
point $t_1$, the second one comes due to $t_2$. Noting, that
\begin{equation}
   P(x_j)=P(x_0)+(-1)^j\,\frac{2\lambda}{27}\, ,
\end{equation}
one has
\begin{eqnarray}
   I&\!\!\!\!\sim \!\!\!\!& 
   e^{iP(x_0)}\frac{3\sqrt{D}}{|b|}
   \int\limits_{-\infty}^{+\infty}\!\ud t\, \left\{
   \exp\left[ i \lambda \left( (-1)^{j}\frac{2}{27}+
   (-1)^{j+1}t^2\right)\right] \right.\nonumber \\
   &\!\!\!\! + \!\!\!\!& \left.
   \exp\left[-i\lambda\left((-1)^{j}\frac{2}{27}+
   (-1)^{j+1}t^2 \right)\right]\right\}
   e^{i\lambda\, \mbox{\footnotesize sgn}(b)\,t^3}.
\end{eqnarray}
It is now obvious that the integrand does not depend on $j$, thus  
\begin{eqnarray}
   I&\!\!\!\!\sim \!\!\!\!& 
   e^{iP(x_0)}\frac{3\sqrt{D}}{|b|}
   \int\limits_{-\infty}^{+\infty}\!\ud t\, \left\{
   \exp\left[ i \lambda \left(\frac{2}{27}-t^2\right)\right] 
   \right.\nonumber \\
   &\!\!\!\! + \!\!\!\!& \left.
   \exp\left[-i\lambda\left(\frac{2}{27}-t^2 \right)\right]\right\}
   e^{i\lambda\, \mbox{\footnotesize sgn}(b)\,t^3}.
\end{eqnarray}
One can see finally that the obtained asymptotical series 
\begin{equation}
   I\sim e^{iP(x_0)}\frac{6\sqrt{D}}{|b|}
   \sqrt{\frac{\pi}{\lambda}}\, \cos\left(
   \frac{2\lambda}{27}-\frac{\pi}{4}\right)+\ldots
\end{equation}
coincides with our previous result (\ref{ralk}).

\subsection*{\normalsize\it 3.5 The perturbation theory approach}
Our asymptotic estimate requires a further explanation in what concerns
the definition of the improper Riemann integral (\ref{I1}). It is  
understood as the limit
\begin{equation}
\label{def1}
   I=\lim_{L\to\infty} I(L), \qquad
   I(L)=\int\limits_{-L}^{L}\ud x\, e^{iP(x)}
\end{equation} 
where we assumed that both stationary points $x_{1,2}$ 
(see eq.(\ref{crit})) belong to the interval $-L<x_{1,2}<L$; 
the asymptotics of $I(L)$ was formed accordingly by two 
independent contributions: $I(L)\sim I_{x_1}(L)+I_{x_2}(L)$.
Integrating in $I_{x_i}(L)$, we took into account that only a 
neighbourhood of a stationary point $x_i$ is important, and 
extended the integration limits to $L\to\infty$. 

Let us suppose now that the coupling $b$ is small, as compared with 
$a$, and one can consider the limit $b\to 0$.  It is easy to see that 
\begin{equation}
   x_1\simeq -\frac{\sigma}{a}+{\cal O}(b), \quad
   x_2\simeq -\frac{2a}{b}+\frac{\sigma}{a}+{\cal O}(b) 
\end{equation}
at small $b$. The first solution is regular at $b\to 0$, but the
second one shows the singular behavior  
\begin{equation} 
   \lim_{b\to 0}|x_2| = +\infty\, .
\end{equation}  
This behavior reflects the increasing importance of the stationary
point $x_1$ in the physical problem for which the cubic term $\sim b$ is
considered as a perturbation. The second stationary point, $x_2$, 
varying as $1/b$, can finally leave the interval $(-L,L)$, changing 
as a result the asymptotics of the integral $I(L)$ to 
\begin{equation}
\label{IL}
   I(L)\sim I_{x_1}(L)\, , \qquad x_2\notin (-L,L).
\end{equation} 

The intuitive argumentation given above must be clarified. One should not
think that the real value of $L$ is very important for the matter.
Actually, one excludes the singular critical point from the phase
$P(x)$ by directly expanding the phase in the neighbourhood of a regular 
stationary point, as we already did it for $I_{x_1}$. What is really 
important here is to conclude that only the regular critical point 
determines the perturbative regime of the system. 
According to this attitude, we use the term ``perturbative regime''
in a wide sense: the standard perturbative expansion in powers of $b$,
which we will obtain below, is merely another way of looking at the
one critical point asymptotics of $I$. 
 
Let us separate the unperturbed part $P_0$ from the perturbation 
$P_\pt$ in the integral $I$
\begin{equation}
   P(x) = P_0(x) + P_\pt (x),\qquad P_0(x)=\sigma x + \frac{a}{2}\, x^2,
   \qquad P_\pt (x)=\frac{b}{6}\, x^3. 
\end{equation}
Expanding the integrand in powers of the coupling $b$, we obtain 
\begin{eqnarray}
   I&\!\!\! =\!\!\!& \int\limits_{-\infty}^{+\infty}\!\ud x\, e^{iP(x)}
   = \int\limits_{-\infty}^{+\infty}\!\ud x\, e^{iP_0(x)}
   \sum_{n=0}^{\infty}\frac{i^n}{n!}P_\pt^n(x)
   \nonumber \\
   &\!\!\! \sim \!\!\!& \sum_{n=0}^{\infty} \frac{i^n}{n!} P_\pt^n
   \left( -i\frac{\partial}{\partial\sigma}\right)
   \int\limits_{-\infty}^{+\infty}\!\ud x\, e^{iP_0(x)}\nonumber \\
   &\!\!\! =\!\!\!& \exp\left\{ \frac{ib}{6} 
   \left( -i\frac{\partial}{\partial\sigma}\right)^3
   \right\}\int\limits_{-\infty}^{+\infty}\!\ud x\, e^{iP_0(x)}
   \equiv I_\pt\, .
\end{eqnarray}
The new feature is the fact that the integrand of $I_\pt$ has only one 
stationary point, $x_\st =-\sigma /a$, as opposed to $I$. The 
singular critical point is gone.

The evaluation of the Gaussian integral is straightforward 
\begin{equation}
   I_0=
   \int\limits_{-\infty}^{+\infty}\!\ud x\, e^{iP_0(x)}
   =\sqrt{\frac{2\pi}{a}} \exp\left( i\left( \frac{\pi}{4}
   -\frac{\sigma^2}{2a} \right)\right), \qquad (a>0).
\end{equation}
This reduces our integral $I_\pt$ to the form
\begin{eqnarray}
   I_\pt &\!\!\! =\!\!\!&\sqrt{\frac{2\pi}{a}}\ e^{i\pi /4}\left[
   1+\frac{ib}{6} \left( -i\frac{d}{d\sigma}\right)^3\!\!
   -\frac{b^2}{72}\left( -i\frac{d}{d\sigma}\right)^6 \right.\nonumber\\
   && \left. -\frac{ib^3}{6^4}\left( -i\frac{d}{d\sigma}\right)^9\!\!
   + \ldots \right] \exp\left(-i\frac{\sigma^2}{2a} \right).
\end{eqnarray}
Up to the terms of order $b^3$ we have now
\begin{eqnarray}
   I_\pt &\!\!\! =\!\!\!& I_0\left[1
   +b  \left(\frac{\sigma}{2a^2}-\frac{i\sigma^3}{6a^3}\right)\! 
   +b^2\left( \frac{5i}{24a^3} + \frac{5\sigma^2}{8a^4} 
       -\frac{5i\sigma^4}{24a^5} - \frac{\sigma^6}{72a^6} \right)
   \right.\nonumber\\
   &&\left.
   +b^3\left( \frac{35i\sigma}{48a^5} +\frac{35\sigma^3}{36a^6} 
   -\frac{7i\sigma^5}{24a^7} - \frac{\sigma^7}{36a^8} 
   +\frac{i\sigma^9}{1296a^9} \right) 
   +{\cal O}(b^4) \right]\, . \nonumber
\end{eqnarray}
Noting, that
\begin{eqnarray}
   \left( 1 + b\beta_1 \!\right. 
   &\!\!\!\! +\!\!\!&\!\left.
   b^2\beta_2 + b^3\beta_3 + \ldots \right) 
   \exp\left[i\left(b\alpha_1 + 
   b^2\alpha_2 +b^3\alpha_3 +\ldots \right)\right] \nonumber\\ 
   &\!\!\!=\!\!\!&
   1 + b(\beta_1 +i\alpha_1 )+b^2 \left(i(\alpha_2 + 
   \alpha_1\beta_1 ) + \beta_2 - \frac{\alpha_1^2}{2}\right)
   \\
   &\!\!\!+\!\!\!&
   b^3\left[ i \left(\alpha_3 +\alpha_1\beta_2 + \alpha_2\beta_1 
                     -\frac{\alpha_1^3}{6} \right)
      +\beta_3 - \alpha_1\alpha_2 - \frac{\beta_1}{2} \alpha_1^2
      \right] + \ldots \nonumber 
\end{eqnarray}
one can represent this result in the more convenient form (up to the same
order of accuracy)
\begin{equation} 
\label{I1pt}
   I_\pt =e^{i\pi /4}\,\sqrt{\frac{2\pi}{a}}\left(1+b\,\frac{\sigma}{2a^2}
   +b^2\,\frac{5\sigma^2}{8a^4} 
   +b^3\,\frac{15\sigma^3}{16a^6}+
   \ldots\right)\ e^{iP_{\eff}(\sigma )},
\end{equation}
where
\begin{equation}
\label{Peff}
   P_{\eff}(\sigma )=-\frac{\sigma^2}{2a}
                    -b\,\frac{\sigma^3}{6a^3}
                    +b^2\left(\frac{5}{24a^3}
                    -\frac{\sigma^4}{8a^5}\right)
                    +b^3\left(\frac{5\sigma}{8a^5}
                    -\frac{\sigma^5}{8a^7}\right)
                    +{\cal O}(b^4).
\end{equation}

The perturbative result $P_{\eff}(\sigma )$ is an approximation, 
which can be systematically improved. In its truncated form the
obtained series differs from the expansion which one can obtain 
from $I_{x_1}$ (see next subsection for details). It is clear, 
however, that if one sums up the series, the perturbative and 
asymptotic estimates will coincide, i.e., $I_\pt =I_{x_1}$.

\subsection*{\normalsize\it 3.6 Resumming the perturbative series}

Let us calculate the contribution to the integral $I$ which comes from 
the critical point $y_1=1$ in eq.(\ref{ias}). It is not difficult to 
find out that exactly this point represents the regular solution at
$b\to 0$. For this part of the integral $I$ we will use the symbol 
$I_{y_1}$, as we did before for $I_{x_i}$, 
\begin{eqnarray}
\label{ias+1}
   I_{y_1}&\!\!\! \sim \!\!\!& 
      e^{iP(x_0)}\left(\frac{2}{|b|}\right)^{1/3}\!\sqrt{|\delta|}
      \int\limits_{-\infty}^{+\infty}\!\ud y\,  
      \exp\left\{ i |\delta |^{3/2} 
      \left(y^2-\frac{2}{3} \right) \right\}
      \nonumber \\
    &\!\!\! \times\!\!\! & 
      \sum_{n=0}^{\infty}\frac{1}{n!}
      \left(i|\delta |^{3/2}\frac{y^3}{3}\right)^n
      \qquad \qquad \qquad (\delta\to -\infty ).
\end{eqnarray}

It is clear that in the last sum only the even powers of $n$ contribute.
For the first two terms ($n=0,2$) one obtains
\begin{equation}
   I_{y_1}\sim \frac{2^{1/3}\sqrt{\pi}}{|b|^{1/3}|\delta|^{1/4}}
   \exp \left\{ i \left[P(x_0) - \frac{2}{3}|\delta |^{3/2} 
   +\frac{\pi}{4}\right]\right\} 
   \left( 1 + \frac{5i}{48|\delta |^{3/2}} +\ldots \right).
\end{equation}
Since
\begin{equation}
   |\delta |^{3/2} = \frac{a^3}{2b^2} 
   \left( 1-\frac{2b\sigma}{a^2}\right)^{3/2},
\end{equation}
we can rewrite this result in the compact form
\begin{equation}
\label{y1}
   I_{y_1} \sim e^{i\pi /4} \sqrt{\frac{2\pi}{a}}\, A(\sigma )\,
   e^{iP'_{\eff}} 
   \left( 1 + \frac{5i}{48|\delta |^{3/2}} +\ldots \right),
\end{equation}
where
\begin{equation}
   A(\sigma ) = 
   \left( 1-\frac{2b\sigma}{a^2}\right)^{-1/4} =
   1 + \frac{b\sigma}{2a^2} + \frac{5b^2\sigma^2}{8a^4} 
   + \frac{15b^3\sigma^3}{16a^6} + 
   {\cal O}(b^4),
\end{equation}
\begin{eqnarray}  
   P'_{\eff}(\sigma )&\!\!\! =\!\!\!&
   -\frac{a\sigma}{b}+\frac{a^3}{3b^2}\left[1-
   \left( 1-\frac{2b\sigma}{a^2}\right)^{3/2}\right] \nonumber\\ 
   &\!\!\! =\!\!\!&
   -\frac{\sigma^2}{2a}-\frac{b\sigma^3}{6a^3}
   -\frac{b^2\sigma^4}{8a^5} - \frac{b^3\sigma^5}{8a^7} +
   {\cal O}(b^4).
\end{eqnarray}
One can explicitly see that the series in powers of $b$ obtained 
from the integral $I_{y_1}$ perfectly coincides\footnote{Actually, we
checked the terms up to and including $b^5$ order.} 
with the perturbative expansions (\ref{I1pt}) and (\ref{Peff}).

The series (\ref{y1}) corresponds to a partial resummation of the 
perturbative series. In order to see what is resummed, 
let us introduce a parameter $t$ according to the substitution: 
$a,b,\sigma\to ta,tb,t\sigma$ in eqs.(\ref{I1pt}) and (\ref{Peff}). 
We find for $I_\pt\to I_\pt(t)$
\begin{equation} 
\label{I1pt-t}
   I_\pt (t) =e^{i\pi /4}\,\sqrt{\frac{2\pi}{a}}\left(1+b\,\frac{\sigma}{2a^2}
   +b^2\,\frac{5\sigma^2}{8a^4} 
   +b^3\,\frac{15\sigma^3}{16a^6}+
   \ldots\right)\ e^{itP_{\eff}(t,\sigma )},
\end{equation}
where
\begin{equation}
\label{Peff-t}
   P_{\eff}(t,\sigma )=-\left(\frac{\sigma^2}{2a}
                               +\frac{b\sigma^3}{6a^3}
                               +\frac{b^2\sigma^4}{8a^5}
                               +\frac{b^3\sigma^5}{8a^7}\right)
                    +\frac{1}{t^2}\left(\frac{5b^2}{24a^3}
                    +\frac{5b^3\sigma}{8a^5}\right)
                    +{\cal O}(b^4).
\end{equation}
It is clear now that the leading term, $P'_{\eff}$, is represented in 
$P_\eff(t)$ by the sum of terms which are $\propto (1/t)^0$, the first 
stationary phase correction is given by terms of order $\propto 1/t^2$, 
and so on. The field theoretical analog of this power series 
expansion in $(1/t)$ is well known as a loop expansion 
\cite{Coleman:1973}.

\vspace{0.7cm}

\section*{\centerline{\large\sc 4. Outlining the problem}}

So far we have worked with a one-dimensional case. We now want
to clarify what actually can be learned from this simple example for 
the functional integral (\ref{intJi}) considered.   

It is clear, for instance, that we cannot follow the approaches discussed in 
subsections 3.1 and 3.2. An analog of eq.(\ref{cpeq}) for our functional 
integral is the equation
\begin{equation} 
\label{sdLr}
   \frac{\partial^2 {\cal L}_r}{\partial R_A \partial R_B} 
   \equiv {\cal L}_{AB}'' =
   G\delta_{AB} + \kappa \Phi_{ABC}R_C = 0
\end{equation}
which cannot be fulfilled.

On the contrary, the system of equations based on the first order 
derivatives  
\begin{equation} 
\label{fdLr}
   \frac{\partial {\cal L}_r}{\partial R_A} =
   GR_A + \Delta_A + \Pi_A + \frac{\kappa}{2}\ \Phi_{ABC}R_BR_C = 0
\end{equation}
is self-consistent and can be solved \cite{Osipov:2004}. Therefore, 
one can obtain the semi-classical asymptotics by analogy with the
stationary phase method discussed in subsection 3.3. It corresponds
to the case without hierarchy, since two critical points are considered. 
We will show, however, that such a system does 
not have a stable vacuum state, i.e., the hierarchy of multi-quark 
interactions is a very important physical requirement for the model. 
Let us discuss the matter in detail here.

\subsection*{\normalsize\it 4.1 Solving equation (\ref{fdLr}) }
We need to recall shortly the solutions of eq.(\ref{fdLr}). Up to some
order in the external mesonic fields, $\Pi_A$, we may write them as a 
polynomial $R_A={\cal R}_A^{(i)}$   
\begin{equation}
\label{Rst}
   {\cal R}_A^{(i)} = H_A^{(i)} 
                    + H_{AB}^{(i)}\Pi_B 
                    + H_{ABC}^{(i)}\Pi_B\Pi_C
                    + H_{ABCD}^{(i)}\Pi_B\Pi_C\Pi_D
                    + \ldots 
\end{equation}
where $i=1,2,\ldots$ denote different possible solutions. The
coefficients $H_{A\ldots}^{(i)}$ depend on $\Delta_a$ and on the coupling 
constants $G,\kappa$, and the higher index coefficients $H_{A\ldots}^{(i)}$ 
are recurrently expressed in terms of the lower ones. For instance, we have
\begin{equation}
   \left(H_{AB}^{(i)}\right)^{-1} = -\left(G\delta_{AB} 
                                    + \kappa\Phi_{ABC}H_C^{(i)}\right), 
\end{equation}
\begin{equation}
   H_{ABC}^{(i)} = \frac{\kappa}{2} \Phi_{DEF} H_{DA}^{(i)} H_{EB}^{(i)} 
                   H_{FC}^{(i)}\ , 
\end{equation}
and so on. 

Putting this expansion in eq.(\ref{fdLr}) one obtains a series of 
self-consistent equations to determine $H_{A\ldots}^{(i)}$. The first 
one is 
\begin{equation}
\label{ha}
   GH_A^{(i)} +\Delta_A+\frac{\kappa}{2}\ \Phi_{ABC}
   H_B^{(i)}H_C^{(i)} = 0.  
\end{equation}

One can always find the trivial solution $H_A=0$, corresponding to the
unbroken vacuum $\Delta_A=0$. There are also non-trivial ones for 
the scalar component, i.e.,
\begin{equation}   
   H_A^{(i)} = (h_a^{(i)},0).
\end{equation}

The number of possible solutions, $i$, depends on the symmetry group. 
The coefficients $h_a^{(i)}$ are determined by the couplings $G, \kappa$
and the mean field value $\Delta_a$. In accordance with the pattern of
explicit symmetry breaking the mean field can have only three non-zero 
components at most with indices $a=0,3,8$. If two of the three indices 
in $A_{abc}$ are $\{0,3,8\}$, then the third one also belongs to this 
set. Thus $\Delta_a$ is the only object which determines the vector 
structure of the solution $h_a^{(i)}$, and therefore $h_a^{(i)}\neq 0$ 
if $a=0,3,8$. It means that in general we have a system of only three 
equations to determine 
$h^{(i)}=h_a^{(i)}\lambda_a=\mbox{diag}(h_u^{(i)},h_d^{(i)},h_s^{(i)})$ 
\begin{equation}
\label{saddle-1}
  \left\{
         \begin{array}{rcl}
\vspace{0.2cm}
        && Gh_u+\Delta_u+\displaystyle\frac{\kappa}{16}h_dh_s=0 \\
\vspace{0.2cm}
        && Gh_d+\Delta_d+\displaystyle\frac{\kappa}{16}h_sh_u=0 \\        
        && Gh_s+\Delta_s+\displaystyle\frac{\kappa}{16}h_uh_d=0\ . \\
         \end{array}
  \right. 
\end{equation}

This system is equivalent to a fifth order equation for a one-type 
variable which can be solved numerically. For two particular cases, when 
$\hat{m}_u=\hat{m}_d=\hat{m}_s$ and $\hat{m}_u=\hat{m}_d\neq\hat{m}_s$, 
eqs.(\ref{saddle-1}) can be solved analytically \cite{Osipov:2004}.
The simplest example: $\hat{m}_u=\hat{m}_d=\hat{m}_s$ (or, equivalently, 
$h_u^{(i)}=h_d^{(i)}=h_s^{(i)}$) corresponds to $SU(3)$ flavor symmetry. 
In this case eq.(\ref{saddle-1}) has two solutions   
\begin{equation}
\label{hsu3}
   h_u^{(1)} = -\frac{8G}{\kappa} \left( 1- 
   \sqrt{1-\frac{\kappa\Delta_u}{4G^2}} \right)\ , \qquad
   h_u^{(2)} = -\frac{8G}{\kappa} \left( 1+ 
   \sqrt{1-\frac{\kappa\Delta_u}{4G^2}} \right).
\end{equation}
If $4G^2>\kappa\Delta$, they are real and will contribute to the 
stationary phase trajectory.  

\subsection*{\normalsize\it 4.2 The lowest order semi-classical asymptotics }
Since the system of equations (\ref{fdLr}) can be solved, we may replace
variables $R_A\rightarrow \bar{R}_A=R_A-{\cal R}_A^{(i)}$ in the 
functional integral (\ref{intJi}) to obtain the semi-classical asymptotics  
\begin{eqnarray}
\label{intJisp}
     {\cal Z}[\Pi ,\Delta ]&\!\!\!\sim \!\!\!& 
     {\cal N} \sum_{i=1}^n\ 
     \exp\left( i\int\ud^4x {\cal L}_\st^{(i)}\right)
     \nonumber \\
     &\!\!\!\times\!\!\!&
     \int\limits^{+\infty}_{-\infty}\prod_A{\cal D}\bar{R}_A\
     \exp\left(\frac{i}{2}\int\ud^4x{\cal L}_{AB}''({\cal R}^{(i)})
     \bar{R}_A \bar{R}_B \right) \nonumber \\
     &\!\!\!\times\!\!\!& \sum_{k=0}^\infty \frac{1}{k!}
     \left(i\frac{\kappa}{3!}\ \Phi_{ABC}\int\ud^4x 
     \bar{R}_A\bar{R}_B\bar{R}_C\right)^k \qquad  (\hbar\to 0)
\end{eqnarray} 
where $n$ is the number of solutions, ${\cal R}^{(i)}_A$, of 
eq.(\ref{fdLr}), ${\cal L}_{AB}''$ has been defined in
eq.(\ref{sdLr}), and
\begin{eqnarray}
\label{Lrsp}
   {\cal L}_\st^{(i)}&\!\!\! =\!\!\!& 
                {\cal R}^{(i)}_A (\Pi_A+\Delta_A) 
                + \frac{G}{2} \left({\cal R}^{(i)}_A\right)^2 
                + \frac{\kappa}{3!}\ \Phi_{ABC}
                {\cal R}^{(i)}_A{\cal R}^{(i)}_B{\cal R}^{(i)}_C
                \nonumber \\
   &\!\!\! =\!\!\!&\frac{G}{6}\left({\cal R}_A^{(i)}\right)^2\!
                +\frac{2}{3}\ {\cal R}^{(i)}_A \left(\Pi_A 
                + \Delta_A\right)
                = h_a^{(i)}\sigma_a + {\cal O}(\Pi^2).
\end{eqnarray}
Here we used eq.(\ref{fdLr}) to eliminate the term proportional to 
$\kappa$. Let us stress that ${\cal L}_\st^{(i)}$ depends on $\kappa$ 
implicitly: $\kappa$ is contained in ${\cal R}^{(i)}_A$, or more precisely 
in the coefficients $H^{(i)}_{A...}$ which are functions of $h^{(i)}_a$. 
This dependence is singular at $\kappa\to 0$. One can see this, for 
instance, from eq.(\ref{hsu3}) where $h_u^{(2)}\to\infty$ for small 
$\kappa$. This behavior reflects the fact that we are far from the
perturbative regime, meaning that the interactions ${\cal L}_\njl$ and 
${\cal L}_\h$ are equally weighted. 

The linear term in the $\sigma$ field is written explicitly. 
This part of the Lagrangian is responsible for the 
dynamical symmetry breaking in the multi-quark system and taken
together with the corresponding part from the Gaussian integration 
over quark fields in eq.(\ref{genf3}) leads us to the gap equation.

At leading order, $k=0$, we have the estimate for eq.(\ref{intJisp})
\begin{equation}
\label{splo}
     {\cal Z}\sim\sum_{i=1}^n A_{(i)} 
     \exp\left( i\int\ud^4x {\cal L}_\st^{(i)}\right) 
     \sim \exp \left(i\int\ud^4x \sum_{i=1}^n 
     h_a^{(i)}\sigma_a + \ldots\right)
\end{equation}
where $A_{(i)}$ is real and proportional to 
\begin{equation}
   A_{(i)} \sim |\det {\cal L}_{AB}''({\cal R}^{(i)})|^{-1/2}.
\end{equation}

Therefore, if one considers the case with 
$\hat{m}_u=\hat{m}_d=\hat{m}_s$, ${\cal Z}$ is given by 
\begin{eqnarray}
\label{gec1}
   {\cal Z}&\!\!\!\sim\!\!\!& 
           \exp\left(i\int\ud^4x \sum_{i=1}^2 
           h_a^{(i)}\sigma_a + \ldots\right) \nonumber \\
           &\!\!\!\sim\!\!\!&\exp\left(-i\frac{8G}{\kappa} 
           \int\ud^4x (\sigma_u + \sigma_d + \sigma_s ) + \ldots\right)
\end{eqnarray}

Let us recall that the quark loop contribution to the gap equation
is well known (see, for instance, \cite{Osipov:2004NPA}). Combining
this known result with the estimate (\ref{gec1}), one can obtain the
corresponding effective potential $U(m)$ as a function of the constituent
quark mass $m$
\begin{eqnarray} 
\label{effpot}
     U(m) &\!\!\! =\!\!\!& \frac{24G}{\kappa}\, m
     -\frac{3N_c}{16\pi^2}\left[m^2\left(\Lambda_q^2 
     - m^2\ln\left(1+\frac{\Lambda_q^2}{m^2}\right)\right)
     \right.\nonumber\\
     &\!\!\! +\!\!\!&\left.
     \Lambda_q^4\ln\left(1+\frac{m^2}{\Lambda_q^2}\right)\right],
\end{eqnarray}
where we consider the case $\hat{m}=0$ for simplicity and $\Lambda_q$
denotes the cutoff of quark loop integrals.

We plot this function in fig.1 for $G/\kappa >0$. 
The parameters are fixed there as $\Lambda_q = 1\,\mbox{GeV}$, 
$G/\kappa = 2.34\cdot 10^{-3}\,\mbox{GeV}^3$. One sees that the 
system has at most a metastable vacuum state\footnote{In the opposite 
case, $G/\kappa <0$, the effective potential does not have extrema 
in the region $m>0$.}.   
We must conclude that the model considered has a fatal flaw and 
can be used only in the framework of the perturbative approach, 
which does assume the hierarchy of multi-quark interactions.

\begin{figure}[ht]
\centerline{\epsfig{file=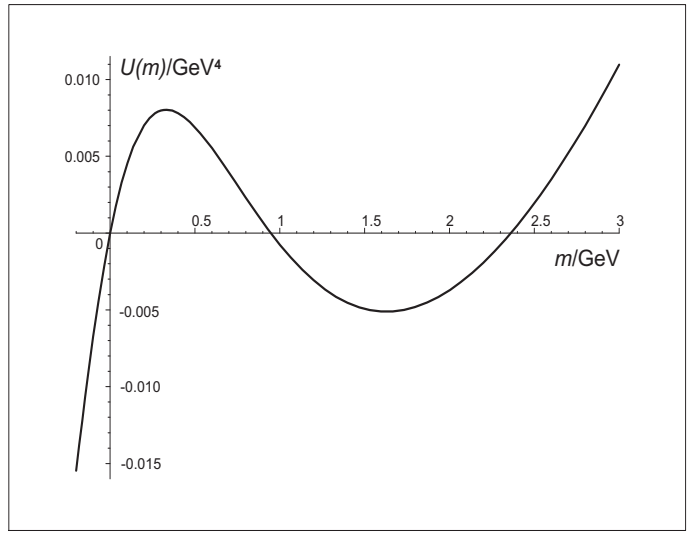,height=7.0cm,width=9cm}}
\caption{The effective potential $U(m)$ (see eq.(\ref{effpot}))
         for the parameter set given in the text   
         as a function of the constituent quark mass $m$. All are given
         in dimensionless units.} 
\label{fig1}
\end{figure}

\vspace{0.7cm}

\section*{\centerline{\large\sc 5. Perturbative expansion of ${\cal Z}$ 
}}

We shall restrict ourselves in this section to the perturbative
treatment of the functional integral (\ref{intJi}). The loop expansion
will be considered in the next section. 

\subsection*{\normalsize\it 5.1 The perturbative series}
Let us devide the Lagrangian (\ref{Lr}) in two parts. The free part, 
${\cal L}_0$, is given by 
\begin{equation}
   {\cal L}_0 (R_A) =  
   \frac{G}{2}\, R_A^2 + R_A(\Pi_A +\Delta_A). 
\end{equation}
The 't Hooft interaction is considered as a perturbation ${\cal L}_I$ 
\begin{equation}
   {\cal L}_I (R_A) =   
   \frac{\kappa}{3!}\ \Phi_{ABC}R_AR_BR_C. 
\end{equation}

Thus the perturbative representation for the functional integral 
(\ref{intJi}) can be written as  
\begin{equation}
\label{Zpt1}
   {\cal Z}= N'\,\exp\left( i\int\! {\cal L}_I (\hat{X}_A) \right)     
   \int\prod_{A} {\cal D} R_A\ e^{i \int\! {\cal L}_0 (R_A)}
\end{equation}
where
\begin{equation}
   \hat{X}_A = -i\,\frac{\delta}{\delta\Pi_A}\, .
\end{equation}

Since the boson fields appear quadratically in eq.(\ref{Zpt1}), they
may be integrated out, yielding 
\begin{equation}
\label{Zpt2}
   {\cal Z}= N\,\exp\left( i\int\! {\cal L}_I (\hat{X}) \right)     
     \exp \left(-i \int
     \frac{\bar{\Pi}_A^2}{2G}\right)
\end{equation}
where $\bar{\Pi}_A=\Pi_A + \Delta_A$. The overall factor 
$N=(-2\pi i/G)^9N'$ is unimportant in the following.   

We want to calculate the effective action $\Gamma_{\eff}$,
which by definition is the phase of ${\cal Z}$
\begin{equation}
\label{Zpt3}
   {\cal Z}=A(\bar\Pi_A )\,\exp \left(i\int\! {\cal L}_{\eff}
   (\bar\Pi_A) \right),
\end{equation}
and $A(\bar\Pi_A )$ is a real function. Comparing (\ref{Zpt2}) and 
(\ref{Zpt3}), one gets
\begin{equation}
\label{Gamma_eff}
   \Gamma_{\eff}=i\ln\frac{A}{N}+\Gamma_0 
   -i\ln\left(1+e^{-i\Gamma_0}
   \left( e^{i\int\! {\cal L}_I}-1\right) e^{i\Gamma_0}\right).
\end{equation}
Here $\Gamma_0$ represents the leading order result for $\Gamma_\eff$
\begin{equation}
\label{gamma0}
   \Gamma_0 = - \frac{1}{2G}\int\bar\Pi_A^2
\end{equation}
while the second logarithm in eq.(\ref{Gamma_eff}) is a source of
$U_\A (1)$ breaking corrections which arise as a series in powers of the
functional derivatives operator 
\begin{equation}
   \hat\Gamma_I = \int\! {\cal L}_I(\hat{X}_A) 
   = \frac{\kappa}{3!}\,\Phi_{ABC}\int \hat{X}_A \hat{X}_B \hat{X}_C\, .
\end{equation}

To make this statement more explicit let us consider the expansion 
\begin{equation}
   \delta = e^{-i\Gamma_0}
   \left( e^{i\hat\Gamma_I}-1 \right) e^{i\Gamma_0} =
   \sum_{m=1}^\infty \frac{i^m}{m!}\left(
   e^{-i\Gamma_0}\hat\Gamma_I e^{i\Gamma_0} \right)^m.
\end{equation}
Taking into account the symmetry properties of the coefficients $\Phi_{ABC}$ 
and our previous result (\ref{contr}), we find 
\begin{eqnarray}
   e^{-i\Gamma_0}\hat\Gamma_I e^{i\Gamma_0} 
   &\!\!\! =\!\!\!& -\frac{\kappa}{3!}\,\Phi_{ABC}\int\left(
      \frac{1}{G^3}\,\bar\Pi_A\bar\Pi_B\bar\Pi_C  
      -\frac{3}{G^2}\,\bar\Pi_A\bar\Pi_B\hat{X}_C 
      \right.\nonumber \\
   &\!\!\! +\!\!\!&\left.\frac{3}{G}\,\bar\Pi_A\hat{X}_B\hat{X}_C
     -\hat{X}_A\hat{X}_B\hat{X}_C \right)
\end{eqnarray}
so that
\begin{equation}
   \delta =\sum_{n=1}^{\infty} \kappa^n\delta_n, 
\end{equation}
represents the effective action (\ref{Gamma_eff}) as a perturbative 
series in powers of $\kappa$. For instance, up to and including the 
second order in $\kappa$ we have  
\begin{equation}
\label{Gamma_eff2}
   \Gamma_{\eff}=i\ln\frac{A}{N}+\Gamma_0 
   -i\kappa\delta_1-i\kappa^2\left(\delta_2-\frac{1}{2}\delta_1^2\right)
   -\ldots
\end{equation}
where
\begin{equation}
\label{delta1}
   \delta_1 = \frac{-i}{3!G^3}\,\Phi_{ABC}\int\bar\Pi_A\bar\Pi_B 
              \bar\Pi_C\, ,
\end{equation}
\begin{eqnarray}
\label{delta2}
   \delta_2 -\frac{\delta^2_1}{2} 
   &\!\!\! =\!\!\!& -\frac{i}{8G^5}\, 
   \Phi_{ABC}\Phi_{AEF}\int \bar\Pi_B \bar\Pi_C \bar\Pi_E \bar\Pi_F 
   \nonumber \\  
   &\!\!\! +\!\!\!& 
   \delta (0) \int \frac{\bar\Pi_A^2}{(8G^2)^2}
   +[\delta (0)]^2 \int \frac{3i}{32G^3}\, .
\end{eqnarray}

The real factor $A(\bar\Pi )$ is always chosen such as to cancel the 
imaginary part of the effective action. To the approximation
considered we have, for instance,
\begin{eqnarray}
\label{A}
   A(\bar\Pi )
   &\!\!\! =\!\!\!& N\,\exp\left\{\delta (0) 
                    \left(\frac{\kappa}{8G^2}\right)^2 \int 
                    \bar\Pi_A^2\right\}\nonumber \\
   &\!\!\! =\!\!\!& N\left(1+ \delta(0) \frac{\kappa^2}{64G^4}
                    \int \bar\Pi_A^2 +\ldots \right).
\end{eqnarray}
It contributes to the measure of the functional integral over 
$\sigma_a , \phi_a$.

\subsection*{\normalsize\it 5.2 The problem with infinities}
The terms with the $\delta (0)$ function require further explanation. 
The fact that auxiliary fields can lead to special problems with
infinities is well-known \cite{Weinberg:1997}. 
For instance, a relevant case can be found in \cite{Bjorken:1965}: 
the new variable $\lambda$ is introduced in the photon self-energy 
diagram in a similar way, i.e., through the integral over a 
$\delta$-function, see eq.(8.18) of the book, which is equal to 1 
and finally, after changes of variables in the whole expression 
eq.(8.19), the logarithmically divergent integral over $\lambda$ 
requires a cutoff.  
     
A case involving a functional integral is given in \cite{Weinberg:2000} 
where the author considers as an example the non-linear $\sigma$-model and 
shows that the Gaussian functional integral with a field dependent 
coefficient in the exponent (there are no derivatives of fields there) 
leads to a $\delta (0)$ factor in the corresponding effective action 
(see also \cite{Abers:1973}). 

In our case auxiliary variables have been introduced by eq.(\ref{1}) 
and changing the order of integration in 
eq.(\ref{genf3}) we finally came to a similar problem with infinities. 

To see the origin of the encountered singularities in detail we shall 
use the language of Feynman diagrams. The graphs contributing to lowest 
order in $\kappa$ are shown in fig.2. 
\begin{figure}[h]
\centerline{\epsfig{file=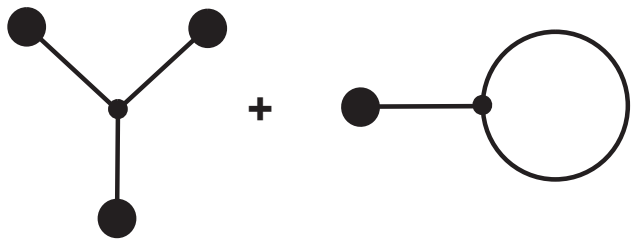,height=3.6cm,width=9cm}}
\caption{The lowest order graphs contributing to $\delta_1$.} 
\label{fig2}
\end{figure}
In these diagrams, a line segment (straight or curved) stands for a
``propagator''
\begin{equation}
\label{propag}
   \Delta_{AB}(x-y)=-(i/G)\delta_{AB}\delta (x-y),
\end{equation} 
extracted from the last exponent in eq.(\ref{Zpt2}). A filled circle
at one end of a line segment corresponds to the external field, 
$i\int\ud^4x\bar\Pi_A(x)$, and a vertex joining three line segments is used
for $i\kappa\Phi_{ABC}\int\ud^4x$.  

The first diagram represents the $\delta_1$ term in eq.(\ref{delta1}).
The contribution of the second tadpole diagram is equal to zero.
Indeed, the vertex contains the group factor $\Phi_{ABC}$. The contraction
of any two indices in this factor by $\delta_{AB}$ from the propagator 
(situation occuring for the tadpole graph) reduces it to zero, 
according to eq.(\ref{contr}). We thus find that tadpole 
diagrams do not contribute due to the flavor structure of the 't Hooft 
interaction. 

To next to leading order in $\kappa$ we have the four graphs shown 
in fig.3. 
\begin{figure}[ht]
\centerline{\epsfig{file=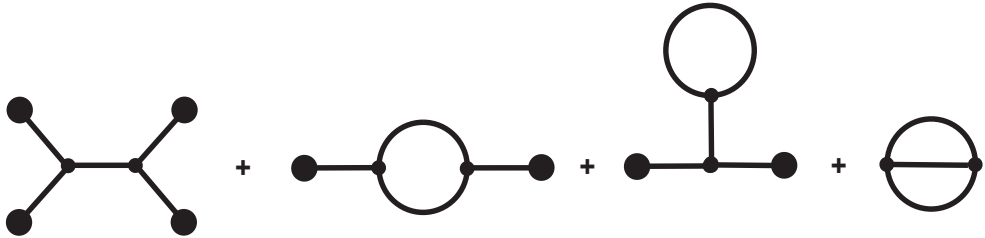,height=3.2cm,width=13cm}}
\caption{The diagrams of the $\kappa^2$ order contributing    
         to $\delta_2$.}
\label{fig3}
\end{figure}
As we just learned, the third diagram does not contribute. The other
ones correspond exactly to the three terms of eq.(\ref{delta2}). The first
tree diagram is finite. The second one-loop diagram has a divergent factor
$\delta (0)$. The last two-loop diagram contributes as $\delta (0)^2$.
These singularities were caused by the local structure of the last 
exponent in eq.(\ref{Zpt2}) or, which is the same, by the $\delta (x-y)$
term in the propagator (\ref{propag}). We believe that if one would start 
from non-local NJL interactions, the singularities could be weaker
or would even disappear.   

The factor $\delta (0)$ requires a regularization. This is an expected
trouble in the NJL model which is nonrenormalizable and, as a consequence, 
the fundamental interactions must be cut off. The cutoff is an effective, 
if crude, implementation of the known short distance behavior of QCD 
within the model. In the next subsection we shall discuss this procedure 
in the framework of the spectral method. The value of the cutoff can 
be fixed by confronting the model with the experiment. In this 
interpretation the model is effectively finite, including the higher 
order corrections.\footnote{ 
Actually, if one works with the NJL model, one must choose among 
several known regularizations. Unfortunately, the dimensional 
regularization (DR) cannot be used. The gap equation in this case does 
not have solutions and as a result there is no dynamical chiral 
symmetry breaking. This is why we cannot simply take advantage of 
the well-known result $\delta (0)=0$ (in DR) to avoid the problem.}

The diagrams are a very convenient language to understand another 
feature related to the multi-loop contributions: the phase factor 
corresponding to the diagram can be simply calculated. Indeed, 
it is easy to see that for any diagram the formula 
\begin{equation}
\label{top1}
   E=2I-3V
\end{equation}
is fulfilled. Here $E$ is the number of external fields, $I$ stands for 
the number of internal lines, and $V$ is the number of vertices.  
On the other hand, the number of loops, $L$, is given by
\begin{equation}
\label{top2}
   L=I-V-E+1\, .
\end{equation}
This is because the number of loops in a diagram is equal to the 
number of $\delta$ functions surviving after all integrations over 
coordinates are performed, exept for one over-all integration related 
to the effective action. Every internal line contributes one $\delta$ 
function, but every vertex or external field carries an integration over
the corresponding coordinate, and thus reduces the number of $\delta$ 
functions by one. 

This result shows that the overall phase factor of a given graph 
$i^{I-E-V}=i^{L-1}$ is entirely determined by the number of loops. 
In particular, diagrams with an even number of loops contribute 
to the effective action (the argument of ${\cal Z}$), while the diagrams 
with odd number of loops contribute to the measure (the modulus of 
${\cal Z}$), see eq.(\ref{Zpt3}).

\subsection*{\normalsize\it 5.3 The spectral representation method}
The problem with $\delta (0)$ singularities can be analysed with more 
accurate methods. In this subsection we shall use the spectral 
representation method to evaluate the integral (\ref{intJ}). 
  
Let us consider, for simplicity, a one-dimensional field theory 
approximation for the integral (\ref{intJ}). The number of 
dimensions does not matter here. The extension to the real case is
straightforward. We also put our system in an interval of size
$L$, i.e. $-L/2\leq x\leq L/2$, assuming the limit $L\to\infty$ in the end 
of calculations. The Fourier decomposition of the fields 
$f_a(x)=\{s_a, p_a, \sigma_a, \phi_a\}$ inside the interval is 
\begin{equation}
\label{fe}
     f_a(x)=\sum_{n=-\infty}^{+\infty} f_n^a\,
            \exp\left(i2\pi \frac{nx}{L}\right).
\end{equation}
This corresponds to the periodic boundary conditions $f_a(-L/2)=f_a(L/2)$. 
The Fourier coefficients $f_n^a$ represent the field $f_a(x)$ inside 
the considered interval   
\begin{equation}
   f_n^a=\frac{1}{L}\int_{-L/2}^{+L/2} \ud x\, f_a(x) 
         \exp\left(-i2\pi \frac{nx}{L}\right), \qquad f_{-n}=f^*_n.
\end{equation} 

Then we have
\begin{equation}
     \int \ud x\, {\cal L}(x) \simeq 
     \int_{-L/2}^{L/2} \ud x\, {\cal L}(x) 
     =L {\cal L}\, ,
\end{equation}
where ${\cal L}$ is devided in two parts: the free one 
${\cal L}_0$, and the perturbation ${\cal L}_I$
\begin{equation} 
   {\cal L}_0=\sum_{n=-\infty}^{+\infty}\left[ 
   \frac{G}{2}\left(s^a_ns^a_{-n}+p^a_np^a_{-n}\right)
   +\sigma^a_n s^a_{-n} + \phi^a_n p^a_{-n}\right], 
\end{equation}
\begin{equation} 
   {\cal L}_I=\frac{\kappa}{32}A_{abc}\sum_{n,m,k}
   s^a_n \left( s^b_{m}s^c_k-3p^b_m p^c_{k}\right)\delta_{n+m+k,0}. 
\end{equation}
 
The functional integral can be understood as the product of integrals
over the Fourier coefficients
\begin{equation}
     \int\limits_{-\infty}^{+\infty} {\cal D}s_a(x) {\cal D}p_a(x) \to
     \prod_{n} \int\limits_{-\infty}^{+\infty} \ud s_n^a \ud p_n^a
\end{equation}
and in the perturbative approximation can be written as
\begin{equation}
     {\cal Z}[\sigma ,\phi ]\sim 
     \exp \left\{iL \hat {\cal L}_I(S_n^a, P_n^a)\right\}
     \prod_{n} \int\limits_{-\infty}^{+\infty} \ud s_n^a \ud p_n^a
     \exp \left(iL {\cal L}_0 \right)
\end{equation} 
where
\begin{equation}
   S^a_n =-\frac{i}{L}\frac{\partial}{\partial \sigma^a_{-n}}\, ,\qquad 
   P^a_n =-\frac{i}{L}\frac{\partial}{\partial \phi^a_{-n}}\, . 
\end{equation}
It is convenient to use the normalized functional ${\cal Z}$ defined as
\begin{equation}
   {\cal Z}=\frac{{\cal Z}[\sigma , \phi ]}{{\cal Z}[0,0]}\, .
\end{equation}

The Gaussian functional integrals can be simply evaluated, for
instance, one has
\begin{eqnarray}
   I[\sigma ]&\!\!\! =\!\!\!&
   \prod_n\int\limits_{-\infty}^{+\infty}\ud s_n\, \exp\left\{
   iL\sum_{n=-\infty}^{+\infty}
   \left(\frac{G}{2}s_ns_{-n}+\sigma_ns_{-n}\right)\right\}\nonumber\\
   &\!\!\! =\!\!\!&\int\limits_{-\infty}^{+\infty}\ud s_0\, \exp\left\{
     iL\left(\frac{G}{2}s_0^2+\sigma_0s_{0}\right)\right\}\nonumber\\
   &\!\!\!\times\!\!\!&\prod_{l=1}^{\infty}\int\limits_{-\infty}^{+\infty}
   \ud s_l \ud s^*_l\, \exp\left\{iL\sum_{l=1}^{+\infty}
   \left(G s_ls^*_{l}+\sigma_ls^*_{l}+\sigma^*_ls_l\right)\right\}
   \nonumber\\
   &\!\!\! =\!\!\!& I[0]\exp\left\{-\frac{iL}{2G}
   \sum_{n=-\infty}^{+\infty} \sigma_n\sigma_{-n}\right\}.
\end{eqnarray}
Thus the functional integration yields
\begin{equation}
   {\cal Z}\sim \exp \left\{iL \hat {\cal L}_I(S_n^a, P_n^a)\right\}
      \exp \left\{-\frac{iL}{2G} \sum_{n=-\infty}^{+\infty} 
      \left(\sigma_n\sigma_{-n} +\phi_n\phi_{-n}\right)\right\}.
\end{equation} 

We should now calculate partial derivatives to obtain the effective
action $\Gamma_{eff}$
\begin{equation}
      {\cal Z}\sim A[\sigma ,\phi ]\, e^{i\Gamma_{eff}}.
\end{equation}  
As before (see subsection 3.2) we have
\begin{equation}
\label{cGamma_eff}
   \Gamma_{eff}=i\ln A[\sigma ,\phi ] +\Gamma_0 
   -i\ln\left(1+e^{-i\Gamma_0}
   \left(e^{i\hat{\Gamma}_I}-1\right)e^{i\Gamma_0}\right),
\end{equation}
where $A[\sigma ,\phi ]$ is fixed by the requirement that the
effective action $\Gamma_{eff}$ is real. Here $\Gamma_0$ represents 
the leading order action
\begin{equation}
\label{cgamma0}
   \Gamma_0 = - \frac{L}{2G}\sum_{n=-\infty}^{\infty}
   \left(\sigma^a_n\sigma^a_{-n}
   +\phi^a_n\phi^a_{-n}\right).
\end{equation}
The logarithm in eq.(\ref{cGamma_eff}) is a source of $U_\A (1)$
breaking corrections which arise as a series in powers of the
partial derivatives  
\begin{eqnarray}
   \hat\Gamma_I &\!\!\! =\!\!\!& L {\cal L}_I(S^a_n,P^a_n)\nonumber \\ 
   &\!\!\! =\!\!\!& \frac{\kappa L}{32}A_{abc}\sum_{n,m,k}
   S^a_n\left(S^b_mS^c_k-3P^b_mP^c_k
   \right)\delta_{n+m+k,0}\, .
\end{eqnarray}

One has the following result 
\begin{eqnarray}
   e^{-i\Gamma_0}\hat\Gamma_I e^{i\Gamma_0} 
   &\!\!\! =\!\!\!& 
      -\frac{\kappa L}{32}\, A_{abc}\sum_{n,m,k}\delta_{n+m+k,0}\left[
      \frac{1}{G^3}\,\sigma^a_n (\sigma^b_m\sigma^c_k -3\phi^b_m\phi^c_k) 
      \right.\nonumber \\
   &\!\!\!-\!\!\!& \frac{3}{G^2}\, (\sigma^b_m\sigma^c_k
                   -\phi^b_m\phi^c_k ) S^a_n 
                   +\frac{6}{G^2}\,\sigma^a_n\phi^b_m P^c_k
   \nonumber\\  
   &\!\!\! +\!\!\!& \frac{3}{G}\,\sigma^a_n (S^b_mS^c_k - P^b_mP^c_k)
           -\frac{6}{G}\,\phi^a_nS^b_mP^c_k \nonumber \\
   &\!\!\! -\!\!\!&\left. S^a_n (S^b_mS^c_k-3P^b_mP^c_k)
   \right].
\end{eqnarray}
Therefore, expanding 
\begin{equation}
   \delta = e^{-i\Gamma_0}
   \left(e^{i\hat{\Gamma}_I}-1\right)e^{i\Gamma_0}
   = \sum_{n=1}^{\infty} \kappa^n\delta_n, 
\end{equation}
one can determine 
\begin{equation}
   \delta_1 = \frac{-iL}{32G^3}\, A_{abc}\sum_{n,m,k}\sigma^a_n 
   (\sigma^b_m\sigma^c_k - 3\phi^b_m\phi^c_k )\delta_{n+m+k,0}\, ,
\end{equation}
\begin{eqnarray}
   \delta_2 &\!\!\! =\!\!\!& \frac{\delta^2_1}{2}
   -i\,\frac{9L}{64(2G)^5}\, A_{abc}A_{a\bar{b}\bar{c}}
   \sum_{m,k,\bar{m},\bar{k}} \delta_{m+k+\bar{m}+\bar{k},0}\left[\, 
   4\sigma^b_m\sigma^{\bar{b}}_{\bar{m}}\phi^c_k\phi^{\bar{c}}_{\bar{k}}
   \right. \nonumber \\
   &\!\!\! +\!\!\!&\left.
   \left(\sigma^b_m\sigma^c_k - \phi^b_m\phi^c_k \right)
   \left(\sigma^{\bar{b}}_{\bar{m}}\sigma^{\bar{c}}_{\bar{k}} 
   - \phi^{\bar{b}}_{\bar{m}} \phi^{\bar{c}}_{\bar{k}} \right)\,\right]
   \nonumber \\  
   &\!\!\! +\!\!\!& 
   \frac{1}{(8G^2)^2}\sum_n \left( \sigma^a_n\sigma^a_{-n} +
   \phi^a_n\phi^a_{-n}\right)\sum_m 1 
   + \frac{3i}{32LG^3}\left(\sum_m 1\right)^2\, ,
\end{eqnarray}
and so forth.

The effective action can be also expanded in $\kappa$
\begin{equation}
   \Gamma_{eff}=i\ln A[\sigma ,\phi ]+\Gamma_0 
   -i\kappa\delta_1 -i\kappa^2\left(\delta_2 -\frac{\delta_1^2}{2}
   \right)+{\cal O}(\kappa^3), 
\end{equation}
where one assumes that $\ln A[\sigma ,\phi ]$ must be also expanded, i.e.,
\begin{eqnarray}
   A[\sigma , \phi ]&\!\!\! =\!\!\!&
   1+\kappa\beta_1 +\kappa^2\beta_2 + {\cal O}(\kappa^3), 
   \nonumber \\ 
   \ln A[\sigma , \phi ]&\!\!\! =\!\!\!& \kappa\beta_1 
   +\kappa^2\left( \beta_2 -\frac{\beta_1^2}{2}\right)+
   {\cal O}(\kappa^3), 
\end{eqnarray}
and, up to the considered order in $\kappa$, we have
\begin{equation}
   \beta_1 =0, \qquad 
   \beta_2 =\frac{1}{(8G^2)^2}\sum_n \left( \sigma^a_n\sigma^a_{-n} +
   \phi^a_n\phi^a_{-n}\right)\sum_m 1\, . 
\end{equation}
Therefore the perturbative action systematically obtains $U_\A (1)$
breaking corrections in $\kappa^n$, which we collect in the
corresponding part of the action, $\Gamma_n$,
\begin{equation}
   \Gamma_{eff} = \sum_{n=0}^{+\infty} \Gamma_n\, ,
\end{equation}
where $\Gamma_0$ is given by eq.(\ref{cgamma0}), and
\begin{eqnarray}
\label{gammas}
   \Gamma_1 &\!\!\! =\!\!\!& -
   \frac{\kappa L}{32G^3}\, A_{abc}\sum_{n,m,k}\sigma^a_n 
   (\sigma^b_m\sigma^c_k - 3\phi^b_m\phi^c_k )\delta_{n+m+k,0}\, ,
   \nonumber\\
   \Gamma_2 &\!\!\! =\!\!\!&
   -\frac{9\kappa^2 L}{64(2G)^5}\, A_{abc}A_{a\bar{b}\bar{c}}
   \sum_{m,k,\bar{m},\bar{k}} \delta_{m+k+\bar{m}+\bar{k},0}\left[\, 
   4\sigma^b_m\sigma^{\bar{b}}_{\bar{m}}\phi^c_k\phi^{\bar{c}}_{\bar{k}}
   \right. \nonumber \\
   &\!\!\! +\!\!\!&\left.
   \left(\sigma^b_m\sigma^c_k - \phi^b_m\phi^c_k \right)
   \left(\sigma^{\bar{b}}_{\bar{m}}\sigma^{\bar{c}}_{\bar{k}} 
   - \phi^{\bar{b}}_{\bar{m}} \phi^{\bar{c}}_{\bar{k}}
   \right)\,\right]. 
\end{eqnarray}
Here an unessential constant (for the physical action) has been omitted 
in $\Gamma_2$.  

Using the well known relation
\begin{equation}
\label{deltasum}
     \sum_{n=-\infty}^{+\infty}\exp\left(i\frac{2\pi x}{L}n\right)
     =L\sum_{n=-\infty}^{\infty}\delta (x-Ln)=L\delta (x)
\end{equation}
where on the last step we used the fact that in the problem considered
all $x$-dependent functions are integrated only in the 
interval $-L/2\leq x\leq L/2$ and, therefore, only the term with 
$n=0$ can contribute, 
one can obtain that
\begin{equation}
\label{s2f}
     \sum_n f^a_nf^b_{k-n}=
     \frac{1}{L}\int_{-L/2}^{L/2} \ud x\, f_a(x) f_b(x) \exp (-i2\pi kx/L).
\end{equation}
As a result, taking the infinite-volume limit $L\to\infty$, we have, 
for instance,
\begin{equation}
   \Gamma_0 = -\frac{1}{2G}\int\limits_{-\infty}^{+\infty} \ud x\,
   \left(\sigma^2_a(x)+\phi^2_a(x)\right)\, .
\end{equation}

Using the following relation 
\begin{equation}
     \sum_{n_{1},n_{2},\ldots ,n_{i}}\!\!\! 
     f^{a_{1}}_{n_{1}} f^{a_{2}}_{n_{2}}\ldots
     f^{a_{i}}_{n_{i}}\, \delta_{n_{1}+n_{2}+\ldots +n_{i},0} 
     =
     \frac{1}{L}\int_{-L/2}^{L/2}\! \ud x\, 
     f^{a_{1}}(x) f^{a_{2}}(x)\ldots
     f^{a_{i}}(x)
\end{equation}
in (\ref{gammas}), and taking the limit $L\to\infty$ one finds that
\begin{eqnarray}
   \Gamma_1 &\!\!\! =\!\!\!& - \frac{\kappa}{32G^3}\, A_{abc}
   \int\limits_{-\infty}^{+\infty}\! \ud x\, \sigma_a(x)\left[ 
   \sigma_b(x)\sigma_c(x) - 3\phi_b(x)\phi_c(x)\right],
   \nonumber\\
   \Gamma_2 &\!\!\! =\!\!\!&
   -\frac{9\kappa^2}{64(2G)^5}\, A_{abc}A_{a\bar{b}\bar{c}}
   \int\limits_{-\infty}^{+\infty}\! \ud x\, \left[\, 
   4\sigma_b(x)\sigma_{\bar{b}}(x)\phi_c(x)\phi_{\bar{c}}(x)
   \right. \nonumber \\
   &\!\!\! +\!\!\!&\left.
   \left(\sigma_b(x)\sigma_c(x) - \phi_b(x)\phi_c(x)\right)
   \left(\sigma_{\bar{b}}(x)\sigma_{\bar{c}}(x) 
   - \phi_{\bar{b}}(x) \phi_{\bar{c}}(x)
   \right)\,\right]. 
\end{eqnarray}

The infinite-volume limit for the expansion coefficient $\beta_2$ 
needs additional explanation. Here one should define carefully the 
limiting procedure. We have
\begin{equation}
\label{beta2}
   \beta_2 =\frac{1}{(8G^2)^2} \int_{-L/2}^{L/2}\! \ud x\, 
   \left( \sigma_a(x)\sigma_a(x) +
   \phi_a(x)\phi_a(x) \right) \frac{1}{L}
   \sum_{m=-\infty}^{+\infty} 1\, . 
\end{equation}
The integral has a smooth $L\rightarrow\infty$ limit, but contains 
the local factor, which can be understood as a $\delta$-function 
singularity,  $\delta (0)$. 
It is clear from eq.(\ref{deltasum}) that the infinite value
\begin{equation}   
   \frac{1}{L}\sum_{n=-\infty}^{+\infty}1
   = \delta (0)  
\end{equation}       
appears as a common factor on the right hand side of eq.(\ref{beta2}), 
representing the density of the Fourier harmonics in the interval. It 
must be regularized by cutting an upper part of the spectrum, e.g.,
\begin{equation}   
     \delta (0)_{\mbox{\footnotesize reg}}
     = \frac{1}{L}\sum_{n=-N}^{N}1 = \frac{2N+1}{L}
\end{equation}       
where $N$ is large enough. Let us recall that the limit $L\to\infty$ 
is to be taken afterwards. Therefore one should clarify the meaning 
of $N$ being large. There are two possibilities. One can fix $N\gg 1$ 
without any relation to the size of the box.
In this case, in the limit $L\to\infty$, the $\delta $-function vanishes. 
Alternatively, one can relate $N$ with $L$ by introducing a 
momentum space cutoff $\Lambda$: $N(L)=L\Lambda /(4\pi)\gg 1$. 
Unlike $L$ the cutoff $\Lambda$ has an obvious physical meaning 
giving the scale of momenta relevant for the problem. 
Indeed, the $n$th harmonic has a momentum $p_n=2\pi n/L$. 
The size of the considered box is the difference 
$p_N-p_{-N}=4\pi N/L=\Lambda$. The scale cannot be eliminated 
by taking the limit $L\to\infty$. One has instead
\begin{equation}
     \delta (0)_{\mbox{\footnotesize reg}}=\frac{\Lambda}{2\pi}
     +\frac{1}{L}\, ,       
\end{equation}     
where only the second term does not contribute in the limit 
$L\rightarrow\infty$. By introducing the cutoff $\Lambda$, we suppose 
that the density of Fourier harmonics has a finite value which can be 
fixed phenomenologically. This is the scenario which we will favor in
the following.

\vspace{0.7cm}

\section*{\centerline{\large\sc 6. The loop expansion of ${\cal Z}$}}

The perturbative series, considered in the previous section, can
be resummed: first summing all diagrams with no closed loops (tree
graphs), then those with one closed loop, etc. As we have already 
discussed, this can be done by generalizing the method of subsec.3.6. 
The tree graphs have been summed in this way in \cite{Reinhardt:1988}, 
and the one-loop graphs in \cite{Osipov:2004}. The two-loop  
approximation is the next to lowest order result which contributes
to the effective action, since the one-loop diagrams contribute solely
to the amplitude.
 
\subsection*{\normalsize\it 6.1 The two-loop approximation}
In this subsection we present a detailed computation of the two-loop 
approximation to the effective mesonic action generated by the functional 
\begin{eqnarray}
\label{Zloop1}
     {\cal Z}[\Pi ,\Delta ]&\!\!\!\sim \!\!\!& 
     {\cal N}  
     \exp\left( i\int\ud^4x {\cal L}_\st^{(i=1)}\right)
     \nonumber \\
     &\!\!\!\times\!\!\!&
     \int\limits^{+\infty}_{-\infty}\prod_A{\cal D}\bar{R}_A\
     \exp\left(\frac{i}{2}\int\ud^4x{\cal L}_{AB}''({\cal R}^{(i=1)})
     \bar{R}_A \bar{R}_B \right) \nonumber \\
     &\!\!\!\times\!\!\!& \sum_{n=0}^\infty \frac{1}{n!}
     \left(i\frac{\kappa}{3!}\ \Phi_{ABC}\int\ud^4x 
     \bar{R}_A\bar{R}_B\bar{R}_C\right)^n \qquad  
\end{eqnarray} 
Note that in comparison with eq.(\ref{intJisp}), only one critical 
point, related to the stable configuration, must be considered. It is 
the solution with $i=1$ in eqs.(\ref{hsu3}).
We shall identify ${\cal L}_\st^{(i=1)}={\cal L}_\st$, and 
${\cal L}_{AB}''({\cal R}^{(i=1)})={\cal L}_{AB}''$ in the following.

By replacing the continuum of spacetime positions with a discrete 
lattice of points surrounded by separate regions of small spacetime
volume $\Omega$, the functional integral (\ref{Zloop1}) may be 
reexpressed as a Gaussian multiple integral over a finite number of
real variables $R_A(x)$ for a fixed spacetime point $x$. We think 
of ${\cal D}R_A$ as the infinite product ${\cal D}R_A\to\prod_x \ud 
R_A(x)$, $\int\ud^4x\to\Omega\sum_x$.
\begin{eqnarray}
\label{Zloop2}
     {\cal Z}[\Pi ,\Delta ]&\!\!\!\sim \!\!\!& 
     {\cal N} \prod_x \left\{\, 
     \exp\left( i \Omega {\cal L}_\st\right) 
     \int\limits^{+\infty}_{-\infty}\prod_A\ud\bar{R}_A\,
     \exp\left(\frac{i}{2}\Omega{\cal L}_{AB}''
     \bar{R}_A \bar{R}_B \right) \right.\nonumber \\
     &\!\!\!\times\!\!\!& \left.\!\!\left(1 -
     \frac{(\kappa\Omega )^2}{2!(3!)^2} 
     \Phi_{ABC}\Phi_{DEF} \bar{R}_A\bar{R}_B\bar{R}_C 
     \bar{R}_D\bar{R}_E\bar{R}_F+\ldots\right)\right\}.   
\end{eqnarray} 
The last sum contains only terms with even powers of $\bar{R}_A$, the
odd powers do not contribute under the Gaussian integration. The dots
mean the terms corresponding to the three-loop contribution and higher.

Let us do the Gaussian integrals 
\begin{equation}
\label{Zloop3}
     {\cal Z}[\Pi ,\Delta ] \sim  
     {\cal N} \prod_x \left\{\, 
     I_0\, \exp\left( i \Omega {\cal L}_\st\right) 
     \left( 1 +i\frac{\kappa^2\Phi_{ABC}\Phi_{DEF}
     \delta_{ABCDEF}}{72\Omega N(N+2)(N+4)}{\cal M}
     +\ldots\right)\right\}. \nonumber\\  
\end{equation} 
Here
\begin{equation}
   {\cal M} =  
   \left(\mbox{tr}\left.{\cal L}''\right.^{-1}\right)^3\!
   + 6\,\mbox{tr}\left.{\cal L}''\right.^{-1}
   \mbox{tr}(\left.{\cal L}''\right.^{-1})^2\!      
   + 8\,\mbox{tr}\left(\left.{\cal L}''\right.^{-1}\right)^3,   
\end{equation}
and $I_0$ is a one-loop contribution 
\begin{equation} 
   I_0=\frac{1}{\sqrt{\det {\cal L}''}}
       \left(\frac{2\pi}{\Omega}\right)^\frac{N}{2}
       \exp\left(i\frac{\pi}{4}\sum\limits_{j=1}^N
       \mbox{sgn}(\lambda_j)\right)
\end{equation}
where $\lambda_j$ are eigenvalues of the $N\times N$ matrix ${\cal
  L}''_{AB}$. In our case $N=18$. $\left.{\cal L}''\right.^{-1}$ is 
an inverse matrix of ${\cal L}''$. The totally symmetric symbol 
$\delta_{ABCDEF}$ generalizes an ordinary Kronecker delta symbol 
$\delta_{AB}$ by the recurrent relation
\begin{equation} 
   \delta_{ABCDEF}= \delta_{AB}\delta_{CDEF}+
                    \delta_{AC}\delta_{BDEF}+\ldots
                   +\delta_{AF}\delta_{BCDE}\, .
\end{equation} 
One can find that
\begin{equation} 
   \Phi_{ABC}\Phi_{DEF}\delta_{ABCDEF} = 6 \Phi_{ABC}\Phi_{ABC}
   = \frac{27}{4}\, .
\end{equation} 

Up to the given accuracy we have in eq.(\ref{Zloop3})
\begin{eqnarray}
\label{Zloop4}
     && 
     \prod_x \left(\, I_0\, e^{i \Omega {\cal L}_\st} 
     \left( 1 +iF \right)\right) = 
     \prod_x \left(\, I_0\, e^{i (\Omega {\cal L}_\st +F)}\right)  
     \nonumber\\
     && 
     = \left(\prod_x I_0 \right) \left( \prod_x
     e^{i (\Omega {\cal L}_\st +F)}\right)
     = \left(\prod_x I_0 \right) 
     e^{i\Omega\sum_x ({\cal L}_\st +F/\Omega )}
\end{eqnarray} 

It shows that in the continuum limit the two-loop correction
contributes to the effective Lagrangian as  
\begin{equation}
\label{2loop}
   {\cal L}_\eff = {\cal L}_\st + 
   \frac{3\kappa^2 [\delta (0)]^2 {\cal M}}{32N(N+2)(N+4)}\, .
\end{equation}   
This is our final expression for the effective Lagrangian in the
two-loop approximation. 

Let us do some estimates to justify the result. For this purpose 
let us simplify the integral (\ref{Zloop1}). After neglecting the 
symmetry group and discretizing the spacetime it takes the form
\begin{equation}
\label{Zls}
     {\cal Z}[\Pi ,\Delta ] \sim \prod_x \int\ud R_x
     \exp\left\{ i \Omega \left( {\cal L}_{\st} +\frac{1}{2}
     {\cal L}_\st''R_x^2 + \frac{1}{3!}{\cal L}_\st'''R_x^3
     \right)\right\}\, .
\end{equation} 
To justify the stationary phase approximation for the integral (\ref{Zls}) 
we assume that 
\begin{equation}
\label{sclc}
     \Omega {\cal L}_\st \gg 1.
\end{equation} 
The dominating role of the Gaussian integral is reflected in the fact that
essential values for $R_x$ in the integral have the order 
$R_x^2\sim 1/(\Omega {\cal L}_\st'')$. For the cubic term it follows then
\begin{equation}
     \Omega {\cal L}_\st'''R_x^3\sim \sqrt{
     \frac{({\cal L}_\st''')^2}{\Omega ({\cal L}_\st'')^3}}
     \sim\sqrt{\frac{\kappa^2}{\Omega G^3}} \sim\sqrt{\zeta}\ ,
\end{equation} 
where we have used that in the model considered here, 
${\cal L}_\st'''\sim\kappa$, 
${\cal L}_\st''\sim G$. If the parameters of the model can be chosen in
such a way that the inequality $\zeta\ll 1$ is fulfilled, the cubic
power of $R_x$ yields terms that go to zero relative to the Gaussian
term as $\zeta\to 0$, and the stationary phase approximation will be 
justified. Note that $\Omega$ may be written as an ultraviolet 
divergent integral regularized by introducing a cutoff $\Lambda$
\begin{equation}
   \Omega^{-1} = \delta^4(0)\sim 
                 \int_{-\Lambda /2}^{\Lambda /2}
                 \frac{\ud^4k_\e}{(2\pi )^4}=
                 \left(\frac{\Lambda}{2\pi}\right)^4.
\end{equation}  
Therefore, the inequality restrics the value of $\Lambda$ from above.

Meanwhile, it is interesting to see that the inequality $\zeta\ll 1$ 
is an exact equivalent of (\ref{sclc}). Indeed, in the essential region, 
i.e., around a sharp minimum, one has ${\cal L}_\st'\sim R_x {\cal L}_\st$, 
${\cal L}_\st''\sim R_x^2{\cal L}_\st$, and so on, thus
\begin{equation}
     \zeta\sim
     \frac{({\cal L}_\st''')^2}{\Omega ({\cal L}_\st'')^3}
     \sim\frac{({R_x^3\cal L}_\st )^2}{\Omega (R_x^2{\cal L}_\st )^3} 
     \sim\frac{1}{\Omega {\cal L}_\st}\ . 
\end{equation} 
 
To summarize, the asymptotical series (\ref{Zloop2}) with the 
ultraviolet cutoff imposed is sensible. One deals here, 
actually, with a series in powers of the dimensionless parameter 
$\zeta$. The expansion is formally justified for $\zeta\ll 1$.

\subsection*{\normalsize\it 6.2 Some final comments}
Several comments are in order:

(1) We believe that it is the first time that loop corrections to the
effective action have been obtained at next to lowest order in 
the bosonization procedure.

(2) This is a nonrenormalizable theory; so we should expect an
explicit dependence on the cutoff in the result. This is indeed 
the case: the $[\delta (0)]^2$ factor is understood as an effective
arbitrary parameter $(\Lambda /(2\pi ))^8$ which together with the other 
coupling constants of the model, $G$ and $\kappa$, forms a 
dimensionless expansion 
parameter $\zeta\sim \kappa^2\Lambda^4/G^3$, the latter must be small 
to justify the loop expansion. This requirement restricts the value of 
$\Lambda$ from above.

(3) The field-dependent factor ${\cal M}$ contains all possible
mesonic vertices, including the $\sigma$-tadpole contribution to
the gap equation, and contributions to the masses of scalar and
pseudoscalar nonets. These contributions might be interesting and 
worth studying phenomenologically.

(4) Without future phenomenological considerations it is difficult
to say at this stage if the loop expansion is a better approximation 
scheme than the ordinary perturbative theory considered above. It 
is certainly not worse if $\kappa$ is small, since the set of graphs 
with $2n$ loops includes, as a subset, all graphs of $\kappa^{n+1}$th 
order or higher in the coupling constant. Thus, in the tree approximation 
the term $\sim\kappa$ will dominate, the two-loop result will include 
the $\kappa$ and $\kappa^2$ order contributions completely, and so on. On 
the other hand, if perturbative corrections are sufficiently large, the 
loop expansion can be more appropriate. 
     
(5) We believe that techniques developed here to obtain the two-loop 
contribution can be easily applied to higher orders in the loop 
expansion. We also hope that some of our findings can be used in
more advanced calculations with non-local effective quark 
Lagrangians.

\vspace{0.7cm}

\section*{\centerline{\large\sc 7. Summary and outlook}}

The purpose of this paper has been to consider several mathematical 
aspects which are related with the $U_\A (1)$ axial symmetry breaking
by the 't Hooft determinant within the framework of the NJL model. Among 
them is the question related to the stability of the ground state, 
the relevance of an hierarchy in multi-quark interactions, and the 
development of techniques for studying the effective Lagrangian beyond 
lowest order in $U_\A (1)$ breaking effects, so that quark -- 
anti-quark bound-state phenomena can be examined in detail.

We have shown that in this picture there is an apparent problem: the 
model has no ground state. This conclusion is based on the stationary 
phase method applied to the generating functional of the model. This 
approach is a well established technique which allows to identify 
straightforwardly all critical points, i.e., all solutions 
${\cal R}_A^{(i)}$ of the stationary phase equations associated with 
the auxiliary bosonic variables, and obtain finally the gap equations
which are the local extrema conditions for the corresponding effective 
potential. It has been shown that these gap equations have no  
phenomenologically acceptable solutions at leading order of the 
stationary phase approach.

If the theory has no ground state, the theory is useless for 
phenomenology. This fact, however, can be used to obtain 
some constraints with a clear phenomenological content. Indeed, 
let us assume for a moment that the 't Hooft coupling $\kappa$ 
scales as $N_c^{1-N_f}$. It would mean that the Lagrangian density 
${\cal L}_\h$ scales with $N_c$, being comparable to ${\cal L}_\njl$. 
We already know that such theory has no ground state and must be 
rejected, thus this counting for $\kappa$ is not acceptable. 
Indeed, all phenomenological facts related with the known solution 
of the $U_\A (1)$ problem provide for abundant evidence in favour of 
a different large $N_c$ identification: $\kappa\sim 1/N_c^{N_f}$. 
   
There are several alternatives to save the situation. We have 
considered here the way based on the perturbative treatment of the 
multi-quark system. Two new results have been obtaind in Sec.3 in 
this connection: 
 
(1) We have shown that the perturbative approach is related to a 
one critical point result of the stationary phase calculations: both
represent the same function, giving different series developments for it.   
   
(2) We have shown that this function, if one would find it, 
is not equal to the starting integral (\ref{I1}), and is also not
an asymptotical series of it. 
   
Both these findings are new to our knowledge and serve to 
understand better the approximations used. They probably cannot be 
rigorously proven for the functional case, but definitely solve the 
problem in the following way: by removing the destructive effect of
the singular critical point in the generating functional, the system
can be treated perturbatively around the stable ground state of the 
NJL model. Therefore, this is a reasonable step for qualitative 
phenomenological considerations at least. 
 
Two series expansions have been discussed in this context. We have 
calculated the first order corrections to both of them and gave the 
complete classification of the terms of the series, separating
contributions to the effective meson Lagrangian from the ones to the 
measure. We have shown that loop corrections serve to obtain 
approximately (i.e., in the framework of the local model which has 
been considered here) the less divergent or even finite contributions 
which would originate within a more refined consideration with
non-local multi-quark vertices.

The new corrections, on which we report, are the direct consequence of
the 't Hooft term in the Lagrangian and in this sense they are 
interesting by themselves and will survive even if theory and 
experiment disagree. Were this the case, it would 
only mean that some essential details in the multi-quark 
effective Lagrangian are still missing, thus stimulating work in 
this direction. In the opposite case, if the predictions are close 
to the data, our findings would put the model on firmer grounds. 

Another commonly used technique is the mean-field method 
\cite{Hatsuda:1994}. It is easy to see that the gap equations that 
one gets within this approach are equivalent to the ones of the 
stationary phase method, when the latter is considered at leading 
order and the contribution of the singular critical point is omitted. 
Therefore, the effective potentials coincide in this case. It is clear 
from this comparison that the loop corrections found in our work can be 
also considered as a step beyond the leading order mean-field result.

There is another way, which yields a rigorous mathematical solution
of the ground state problem above: one can take into account the 
eight-quark or higher order interactions with hope to stabilize 
the vacuum. This approach will be considered elsewhere \cite{Osipov:2005}.

\section*{Acknowledgements}
This work has been supported by grants provided by Funda\c c\~ao para
a Ci\^encia e a Tecnologia, POCTI/FNU/50336/2003 and Centro de 
F\ii sica Te\'orica unit 535. This research is part of the EU 
integrated infrastructure initiative Hadron Physics project under 
contract No.RII3-CT-2004-506078. A.A. Osipov also gratefully
acknowledges the Funda\ca o Calouste Gulbenkian for financial support.  
B. Hiller and A.A. Osipov are very grateful to V. Miransky 
and V.N. Pervushin for discussions.


\section*{\centerline{\large\sc Appendix }}


\subsection*{\normalsize\it The Airy function}

The Airy function $Ai (x)$ is defined as 
\begin{equation}
   Ai (x) = \frac{1}{2\pi}
            \int\limits_{-\infty}^{+\infty}\! \ud t\,
       \exp\left( i \left( \frac{t^3}{3}+ x t\right)\right)\, ,
\end{equation}
at real values of $x$ and can be analytically continued to the whole
complex plane $x\to z$ as an entire function of $z$. \\ 
(a) It is real, if $z$ is real.\\
(b) It decreases exponentially for $|\mbox{arg}\, z|<\pi /3$.\\
(c) It increases exponentially for 
    $\pi /3<\mbox{arg}\, z<\pi$, and
    $-\pi <\mbox{arg}\, z<-\pi /3$.\\
(d) It oscillates at $\mbox{arg}\, z=\pm\pi/3, \pi$.\\
(e) On the real axis the function has the following asymptotics  
\begin{equation}
\label{Airey-as1}
   Ai (x)\sim \frac{x^{-1/4}}{2\sqrt{\pi}}\exp 
         \left(-\frac{2}{3}x^{3/2}\right)\qquad (x\to +\infty ).
\end{equation} 
\begin{equation}
\label{Airey-as2}
   Ai (x)\sim \frac{(-x)^{-1/4}}{\sqrt{\pi}}\left[ 
         \cos\left(\frac{2}{3}(-x)^{3/2}-\frac{\pi}{4}\right)
         +{\cal O}(x^{-3/2})\right]\qquad (x\to -\infty ).
\end{equation}



\begin{thebibliography}{99}

\bibitem{Weinberg:1979} S. Weinberg, Physica A96 (1979) 327.
\bibitem{Gasser:1983} J. Gasser and H. Leutwyler, Ann. of Phys. (N.Y.)
      158 (1984) 142; J. Gasser and H. Leutwyler, Nucl. Phys. B250
      (1985) 465.
\bibitem{Diakonov:2003} see, for instance, recent reviews:
      D. Diakonov, Progr. in Part. and Nucl. Phys. 51 (2003) 173,
      hep-ph/0212026;
      E. V. Shuryak, Phys. Reports 391 (2004) 381. 
\bibitem{Hooft:1978}   
      G. 't Hooft, Phys. Rev. D14 (1976) 3432; 
      Erratum: {\it ibid} D18 (1978) 2199.
\bibitem{Simonov:1997} Yu. A. Simonov, Phys. Lett. B412 (1997) 371,
      hep-th/9703205.
\bibitem{Bali:2001} G. S. Bali, Phys. Rept. 343 (2001) 1,
      hep-ph/0001312.
\bibitem{Simonov:2002} Yu. A. Simonov, Phys. Rev. D65 (2002) 094018,
      hep-ph/0201170.
\bibitem{Nambu:1961} 
      Y. Nambu and G. Jona-Lasinio, Phys. Rev. 122 (1961) 345; 
      124 (1961) 246;
      V. G. Vaks and A. I. Larkin, Zh. \'{E}ksp. Teor. Fiz. 40 
      (1961) 282.  
\bibitem{Eguchi:1976} T. Eguchi, Phys. Rev. D14 (1976) 2755;
      K. Kikkawa, Progr. Theor. Phys. 56 (1976) 947.
\bibitem{Volkov:1982} M. K. Volkov and D. Ebert, Sov. J. Nucl. Phys.
      36 (1982) 736;
      D. Ebert and M. K. Volkov Z. Phys. C16 (1983) 205. 
\bibitem{Ebert:1986} 
      M. K. Volkov, Ann. Phys. (N.Y.) 157 (1984) 282;
      A. Dhar and S. Wadia, Phys. Rev. Lett. 52 (1984) 959;
      A. Dhar, R. Shankar and S. Wadia, Phys. Rev. D31 (1985) 3256;
      D. Ebert and H. Reinhardt, Nucl. Phys. B271 (1986) 188;
      C. Sch\"uren, E. R. Arriola and K. Goeke, Nucl. Phys. A547
      (1992) 612;
      J. Bijnens, C. Bruno and E. de Rafael, Nucl. Phys. B390 (1993)
      501, hep-ph/9206236;
      V. Bernard, A. A. Osipov and U.-G Mei\ss ner, Phys. Lett. B324
      (1994) 201, hep-ph/9312203;
      V. Bernard, A. H. Blin, B. Hiller, Yu. P. Ivanov, A. A. Osipov
      and U.-G Mei\ss ner, Ann. Phys. (N.Y.) 249 (1996) 499, 
      hep-ph/9506309;
      J. Bijnens, Phys. Rep. 265 (1996) 369, hep-ph/9502335.
\bibitem{Dorokhov:1992} A. E. Dorokhov, Y. A. Zubov and
      N. I. Kochelev, Sov. J. Part. Nucl. 23 (1992) 522;  
      E. V. Shuryak, Rev. Mod. Phys. 65 (1993) 1.
\bibitem{Okubo:1963} 
      S. Okubo, Phys. Lett. B5 (1963) 165;
      G. Zweig, CERN Report No. 8419/TH412 (1964);
      I. Iizuka, Progr. Theor. Phys. Suppl., 37 (1966) 38; 21.
\bibitem{Bernard:1988} 
      V. Bernard, R. L. Jaffe and U.-G. Meissner, Phys. Lett. B198 
      (1987) 92;   
      V. Bernard, R. L. Jaffe and U.-G. Meissner, Nucl. Phys. B308 
      (1988) 753. 
\bibitem{Reinhardt:1988} 
      H. Reinhardt and R. Alkofer, Phys. Lett. B207 (1988) 482.
\bibitem{Hatsuda:1994}
      S. P. Klevansky, Rev. Mod. Phys. 64 (1992) 649;
      T. Hatsuda and T. Kunihiro, Phys. Rep. 247 (1994) 221,
      hep-ph/9401310. 
\bibitem{Witten:1979a} 
      E. Witten, Nucl. Phys. B156 (1979) 269;
      G. Veneziano, Nucl. Phys. B159 (1979) 213.
\bibitem{Diakonov:1996} 
      D. Diakonov, {\it Chiral symmetry breaking by instantons},
      Lectures at the Enrico Fermi school in Physics, Varenna,
      June 27 -- July 7, 1995, hep-ph/9602375;
      D. Diakonov, {\it Chiral quark-soliton model}, Lectures at the
      Advanced Summer School on Non-Perturbative Field Theory,
      Peniscola, Spain, June 2-6, 1997, hep-ph/9802298. 
\bibitem{Diakonov:1985} D. Diakonov and V. Petrov, Sov. Phys. JETP 62
      (1985) 204, 431; Nucl. Phys. B272 (1986) 457;
      D. Diakonov and V. Petrov, Leningrad preprint 1153 (1986); 
      D. Diakonov, V. Petrov and P. Pobylitsa, Nucl. Phys. B306 (1988) 
      809.
\bibitem{Osipov:2002} 
      A. A. Osipov and B. Hiller, Phys. Lett. B539 (2002) 76, hep-ph/0204182.
\bibitem{Osipov:2004}
      A. A. Osipov and B. Hiller, Eur. Phys. J. C35 (2004) 223,
      hep-th/0307035.  
\bibitem{Coleman:1973} S. Coleman and E. Weinberg, Phys. Rev. D7
      (1973) 1888.
\bibitem{Osipov:2004NPA} A. A. Osipov and B. Hiller, Phys. Rev. D
      62 (2000) 114013, hep-ph/0007102;
      A. A. Osipov, H. Hansen and B. Hiller,
      Nucl. Phys. A745 (2004) 81, hep-ph/0406112.
\bibitem{Weinberg:1997} S. Weinberg, Phys. Rev. D56 (1997) 2303, 
      hep-th/9706042.
\bibitem{Bjorken:1965} J. Bjorken and S. Drell, 
      ``Relativistic Quantum Mechanics'', McGraw-Hill, New York, 1964. 
      (see in Chapter VIII, Section 8.2). 
\bibitem{Weinberg:2000} S. Weinberg, ``The Quantum
      Theory of Fields'', Volume 1, Cambridge University Press, 1995 
      (see in Chapter IX, in the end of \S 9.3 ). 
\bibitem{Abers:1973} 
      E. S. Abers and B. W. Lee, Phys. Rep. Vol.9, 
      (1973) 1 (see eq.(11.20) there).
\bibitem{Osipov:2005} A. A. Osipov, B. Hiller and J. da Provid\^encia,
      hep-ph/0508058.
\end{thebibliography}
\end{document}